\documentclass[twocolumn]{aastex63}

\newcommand {\msun}{\,M$_\odot \ $}
\newcommand {\dg}{$^\circ$}
\newcommand {\kms}{$\rm{km\ s^{-1}}$}

\received{September 8, 2021}
\revised{September 24, 2021}
\accepted{September 24, 2021}
\submitjournal{ApJ}

\shorttitle{The ejecta of SNR 0540 in 3D}
\shortauthors{Larsson et al.}

\begin{document}

\title{Clumps and rings of ejecta in SNR~0540--69.3 as seen in 3D}

\correspondingauthor{J. Larsson}
\email{josla@kth.se}

\author[0000-0003-0065-2933]{J. Larsson}
\affiliation{Department of Physics, KTH Royal Institute of Technology, The Oskar Klein Centre, AlbaNova, SE-106 91 Stockholm, Sweden}

\author[0000-0003-1546-6615]{J. Sollerman}
\affiliation{Department of Astronomy, Stockholm University, The Oskar Klein Centre, AlbaNova, SE-106 91 Stockholm, Sweden}

\author[0000-0002-3464-0642]{J. D. Lyman}
\affiliation{Department of Physics, University of Warwick, Gibbet Hill Road, Coventry CV4 7AL, UK}

\author[0000-0001-6815-4055]{J. Spyromilio}
\affiliation{European Southern Observatory, Karl-Schwarzschild-Strasse 2, D-85748 Garching, Germany}

\author[0000-0002-7746-8512]{L. Tenhu}
\affiliation{Department of Physics, KTH Royal Institute of Technology, The Oskar Klein Centre, AlbaNova, SE-106 91 Stockholm, Sweden}

\author[0000-0001-8532-3594]{C. Fransson}
\affiliation{Department of Astronomy, Stockholm University, The Oskar Klein Centre, AlbaNova, SE-106 91 Stockholm, Sweden}

\author[0000-0002-3664-8082]{P. Lundqvist}
\affiliation{Department of Astronomy, Stockholm University, The Oskar Klein Centre, AlbaNova, SE-106 91 Stockholm, Sweden}

\begin{abstract}
The distribution of ejecta in young supernova remnants offers a powerful observational probe of their explosions and progenitors. Here we present a 3D reconstruction of the ejecta in SNR~0540-69.3, which is an O-rich remnant with a pulsar wind nebula located in the LMC. We use observations from VLT/MUSE to study H$\beta$, [\ion{O}{3}]~$\lambda \lambda 4959, 5007$, H$\alpha$, [\ion{S}{2}]~$\lambda \lambda 6717, 6731$,  [\ion{Ar}{3}]~$\lambda 7136$  and [\ion{S}{3}]~$\lambda 9069$. This is complemented by 2D spectra from VLT/X-shooter, which also cover  [\ion{O}{2}]~$\lambda \lambda 3726, 3729$ and [\ion{Fe}{2}]~$\lambda 12567$. We identify three main emission components: (i) Clumpy rings in the inner nebula ($\lesssim 1000$~\kms) with similar morphologies in all lines; (ii) Faint extended [\ion{O}{3}] emission dominated by an irregular ring-like structure with radius $\sim 1600$~\kms\ and inclination $\sim 40$\dg, but with maximal velocities reaching $\sim 3000$~\kms; and (iii) A blob of H$\alpha$ and H$\beta$  located southeast of the pulsar at velocities  $\sim 1500-3500$~\kms. We analyze the geometry using a clump-finding algorithm and use the clumps in the  [\ion{O}{3}] ring to estimate an age of $1146 \pm 116$~years. The observations favor an interpretation of the [\ion{O}{3}] ring as ejecta, while the origin of the H-blob is more uncertain. An alternative explanation is that it is the blown-off envelope of a binary companion. From the detection of Balmer lines in the innermost ejecta we confirm that SNR~0540 was a Type II supernova and that hydrogen was mixed down to low velocities in the explosion. 

\end{abstract}


\section{Introduction} 
\label{sec:intro}

Modern three-dimensional supernova (SN) explosion simulations demonstrate the importance of hydrodynamical instabilities for the successful launch of the shock, which results in asymmetric ejecta with signatures of mixing. Predictions include fast Ni plumes and anisotropic fallback \cite[e.g.,][]{Stockinger2020,Gabler2021,Sandoval2021}. The fact that SNe are not spherical is known from a large number of observations, not least from imaging of young SN remnants \citep[SNR, see][for a review]{Milisavljevic2017}. In these objects, the observed Doppler shifts can be directly translated to distance from the center of explosion along the line of sight, assuming a free expansion of the ejecta. This means that the 3D emissivity of the ejecta can be directly reconstructed from the combination of spectral and imaging information. 

SNR~0540-69.3  (hereafter SNR~0540) is a $\sim 1000$ year old SNR in the Large Magellanic Cloud (LMC). It is often referred to as the ``Crab twin" as its 50~ms pulsar and associated pulsar wind nebula (PWN) are similar to those of the Crab Nebula. However, many other properties of the remnant are not Crab-like. Most notably,  SNR~0540 is O-rich, is thought to originate from a more massive progenitor ($\gtrsim 15~M_{\sun}$, \citealt{Chevalier2006,Williams2008}) and shows a shell of emission due to the interaction between the blast wave and surrounding medium. The shell has a radius of $\sim 30^{\prime\prime}$ and emits in radio, millimeter and X-rays \citep{Manchester1993,Hwang2001,Brantseg2014,Lundqvist2020}. 

The optical line emission in SNR~0540 is powered by the PWN. Most of the emission is concentrated to a region of $\sim 4^{\prime\prime}$ radius \citep{Kirshner1989,Serafimovich2005,Morse2006,Williams2008},\footnote{A more detailed analysis of the remnant than that in \cite{Serafimovich2005} can be found in \cite{Lundqvist2021}.} but there is also a faint halo of [\ion{O}{3}] emission that extends to at least $\sim 8^{\prime\prime}$ \citep{Morse2006}. The detection of H lines from the inner region \citep{Serafimovich2005,Morse2006} implies that the SN was a Type II, something that was previously not appreciated.  

The  [\ion{O}{3}]~$\lambda 5007$ and [\ion{S}{2}]~$\lambda \lambda 6717, 6731$ emission from the inner region have previously been studied in 3D using VLT/VIMOS observations, which revealed ring-like structures \citep{Sandin2013}. Here we present new information about the 3D structure of the ejecta using observations from VLT/MUSE and X-shooter. The analysis includes a number of lines not previously studied in 3D, as well as higher-resolution maps for [\ion{S}{2}] and [\ion{O}{3}] for a larger field of view (FOV) than the VIMOS maps of \cite{Sandin2013}.  We focus on emission that can be clearly connected to fast-moving material associated with SNR~0540, which turns out to be confined to a region within $\sim 17^{\prime \prime}$ of the pulsar. The outer regions, including any possible emission associated with the radio/X-ray shell, will be studied as part of a series of forthcoming papers based on the MUSE and X-shooter observations of SNR~0540.

This paper is organized as follows. In Section~\ref{sec:obs} we describe the observations and data reduction. Section~\ref{sec:methods} details how we construct the 3D maps from the observations, while the results are presented in Section~\ref{sec:results}, where we use the age as derived in Section~\ref{sec:results-analysis}. Finally, we discuss the results in Section~\ref{sec:disc} and summarize our conclusions in Section~\ref{sec:conclusions}.

\section{Observations and data reduction}
\label{sec:obs}

Multi Unit Spectroscopic Explorer (MUSE; \citealt{Bacon2010}) observations of SNR 0540 were taken as part of ESO program 0102.D-0769, utilizing the wide-field mode adaptive optics (WFM-AO). The instrument was used in extended mode, which provides a slightly bluer wavelength cutoff, allowing for the capture of H$\beta$. Observations were split into four observing blocks between January and March 2019, with each block consisting of $3\times770$~s exposures.  
A 230~s exposure was taken immediately after each block at an offset of a few arcminutes for the purposes of sky subtraction. The final science exposure of the second observing block was not used as image quality deteriorated during this time, resulting in a total exposure time of $\sim 2.35$~h. The reduced data were retrieved from ESO Phase 3 data products, having been reduced with MUSE reduction pipeline version 2.6.2. The MUSE WFM-AO observations provide integral-field spectroscopic observations covering a $1^{\prime} \times 1^{\prime}$ FOV at a spatial resolution of $0\farcs{3}-0\farcs{4}$ with $0\farcs{2}$  sampling. The observations cover the wavelength range 4650--9300~\AA\ (with a gap between 5760--6010~\AA\ due to the Na laser) at a spectral resolution of 1750--3750.

X-shooter observations of SNR~0540 were performed between 23 October and 2 November 2019 as a part of ESO program 0104.D-0532. The instrument has three arms (UVB, VIS and NIR), which together cover the wavelength interval $3000-25,000$~\AA.  We used slit widths of $1\farcs{6}$ (UVB), $1\farcs{5}$ (VIS) and $1\farcs{2}$ (NIR), which provide a spectral resolution of 3200, 5000 and 4300, respectively. The orientation of the slit is shown in Figure~\ref{fig:muse_images_inner}. The seeing during the observations was $0\farcs{8}-1^{\prime \prime}$ and the airmass was 1.5. 

The observations were performed as a nodding pattern in six observing blocks, with each block containing four exposures of 280, 190 and  300~s in UVB, VIS and NIR, respectively. SNR~0540 is too large to fit within twice the $11^{\prime\prime}$ slit length. Thus, although we nodded along the slit, the data were not reduced as a nodding sequence, as that would have resulted in subtraction of flux from the outer regions of the remnant. The data were instead reduced in STARE mode using the standard ESO pipeline, followed by shifting and adding the individual STARE observations. 
Sky subtraction was undertaken using the flux at the edges of the slits in the combined frame. We also checked that individual frames could be sky subtracted followed by the combination with minor differences from the adopted sequence described above. In the UVB and VIS, we did not perform sky subtraction of the narrow lines from the interstellar medium (ISM) that overlap with the broad ejecta lines. As described in Section~\ref{sec:methods}, we were able to remove these lines more accurately by fitting the spectra at each position along the slit. The normal nodding observing sequence would automatically compensate for bad pixels while simultaneously sky-subtracting. In our case, and for the NIR arm in particular, bad pixels were identified and corrected by interpolation using adjacent pixels before combining the frames.

\section{Construction of emissivity maps}
\label{sec:methods}

We analyze the strongest emission lines to determine the 3D structure of the ejecta. These are  \hbox{[\ion{O}{3}]}~$\lambda \lambda 4959, 5007$, H$\alpha$, \hbox{[\ion{S}{2}]}~$\lambda \lambda 6717, 6731$,  [\ion{Ar}{3}]~$\lambda 7136$  and [\ion{S}{3}]~$\lambda 9069$. In the X-shooter data we also include [\ion{O}{2}]~$\lambda \lambda 3726, 3729$ and [\ion{Fe}{2}]~$\lambda 12567$, which are outside the MUSE wavelength range. In addition to these lines, we analyze H$\beta$ in both data sets. While this line has a relatively low signal-to-noise ratio (S/N), it provides important information about the ejecta and also help us interpret the H$\alpha$ emission, which suffers from blending with strong emission from [\ion{N}{2}]~$\lambda \lambda 6548, 6584$. We focus on the morphology of the different lines and do not correct the spectra for reddening. 
 
SNR~0540 is surrounded by a complex ISM that produces narrow emission lines. The centroid velocity and width of these lines vary with position and line species. In the following we correct all spectra for the average systemic velocity of the ISM near the center of the remnant, which we determine to be $277$~\kms\ in both data sets independently. The standard deviation about the average velocity is 1.5 and 6.5~\kms\ for the dozen different lines fitted in X-shooter and MUSE, respectively. 

We also take $277$~\kms\  as the center of explosion along the line of sight when creating 3D maps. In the image plane, we take the pulsar position as the center of the explosion. This is a reasonable approximation given the upper limit on the transverse velocity of the pulsar of $<250$~\kms\ \citep{Mignani2010}. We translate distances from the pulsar to ejecta velocities assuming a free expansion and an age of 1100 years (Section \ref{sec:results-age}). For a distance to the LMC of 49.6~kpc \citep{Pietrzynski2019},  this age implies that $1''$ (0.24~pc) corresponds to $214$~\kms. 

In order to isolate the emission from the ejecta, it is necessary to subtract the continuum emission from the pulsar, PWN and nearby stars, as well as the narrow line emission from the ISM. For the MUSE data cube, we perform the subtraction based on fits to the spectra in each spatial pixel.  First, the continuum level was determined by fitting straight lines to $\sim 1000$~\kms\ intervals on both sides of each line. The contribution from the ISM was then determined by fitting a Gaussian plus a straight line to the central part of the continuum-subtracted spectra (between $\sim \pm 130-250$~\kms\ for the different lines, considering the wavelength-dependence of the spectral resolution). 

The reason for removing ISM lines by fitting is the significant variations of these lines across the remnant, which implies that simply subtracting a background spectrum usually leaves large residuals. In pixels where the fit to the ISM lines failed or returned unreasonable values, we used fit parameters obtained from background spectra. In these cases, we fixed the centroid and width of the Gaussian to the background values and adjusted only the normalization to subtract the line. The approximation of the ejecta profile by a straight line over the narrow velocity intervals used in these fits is good in most cases. However, it clearly introduces a systematic uncertainty that should be kept in mind when interpreting the results at low Doppler shifts. 

The wavelength intervals covered by some ejecta lines also include other weak ISM lines, such as \ion{He}{1}~$\lambda5016$ overlapping with the [\ion{O}{3}] profile. These were removed in a similar way as described above. The data cubes were carefully inspected after the subtractions to identify any residuals not associated with ejecta. This revealed some signs of ISM interaction (including the western ``F1" region studied in \citealt{Serafimovich2005}) and problems with a small number of stars. While most stellar spectra are reasonably well described by straight lines over the narrow wavelength ranges considered here, there are exceptions, such as stars exhibiting strong absorption in H$\beta$. Problematic ISM regions and stars were masked out or dealt with in a custom manner (e.g., by adjusting the wavelength interval of the continuum fits or applying sigma-clipping in the region of some stars). We verified that there was no confusion with significant signal from the ejecta when masking out problematic regions.

The [\ion{O}{3}] doublet was deblended using the theoretical ratio  $I(\rm {[O\,III]}~\lambda 5007)/I(\rm{ [O\,III]}~\lambda 4959\rm )~=~3$. For the [\ion{S}{2}] emission, we take the ratio  $I({\rm [S\,II] }~\lambda 6731) / I({\rm [S\,II]}~\lambda 6716)~=~1.0$, which is close to the average number of \cite{Serafimovich2005} and \cite{Sandin2013}. We will investigate possible spatial variations of this line ratio in future work. We do not attempt to deblend the H$\alpha + $[\ion{N}{2}] complex, but note that the overlap between these lines in wavelength space is relatively small in the X-shooter spectra. Finally, for 3D visualization purposes, we remap the cleaned cubes for all lines to a uniform grid with a spacing of $40$~\kms.

For X-shooter data, we fit and subtract the continuum in each row of the 2D spectra, analogous to the procedure used for MUSE. The narrow ISM lines in the UVB and VIS were also removed as for MUSE, using fits with Gaussians. This method leaves clear residuals inside $\sim \pm 150$~\kms, but nevertheless produces significantly better results than simply subtracting background spectra from the edges of the slits. On the other hand, the narrow lines in the NIR were best removed using standard background subtraction. In the NIR we also removed some remaining bad pixels in the [\ion{Fe}{2}] wavelength region using a sigma-clipping routine.

\section{Results}
\label{sec:results}

\begin{figure*}[t]
\plotone{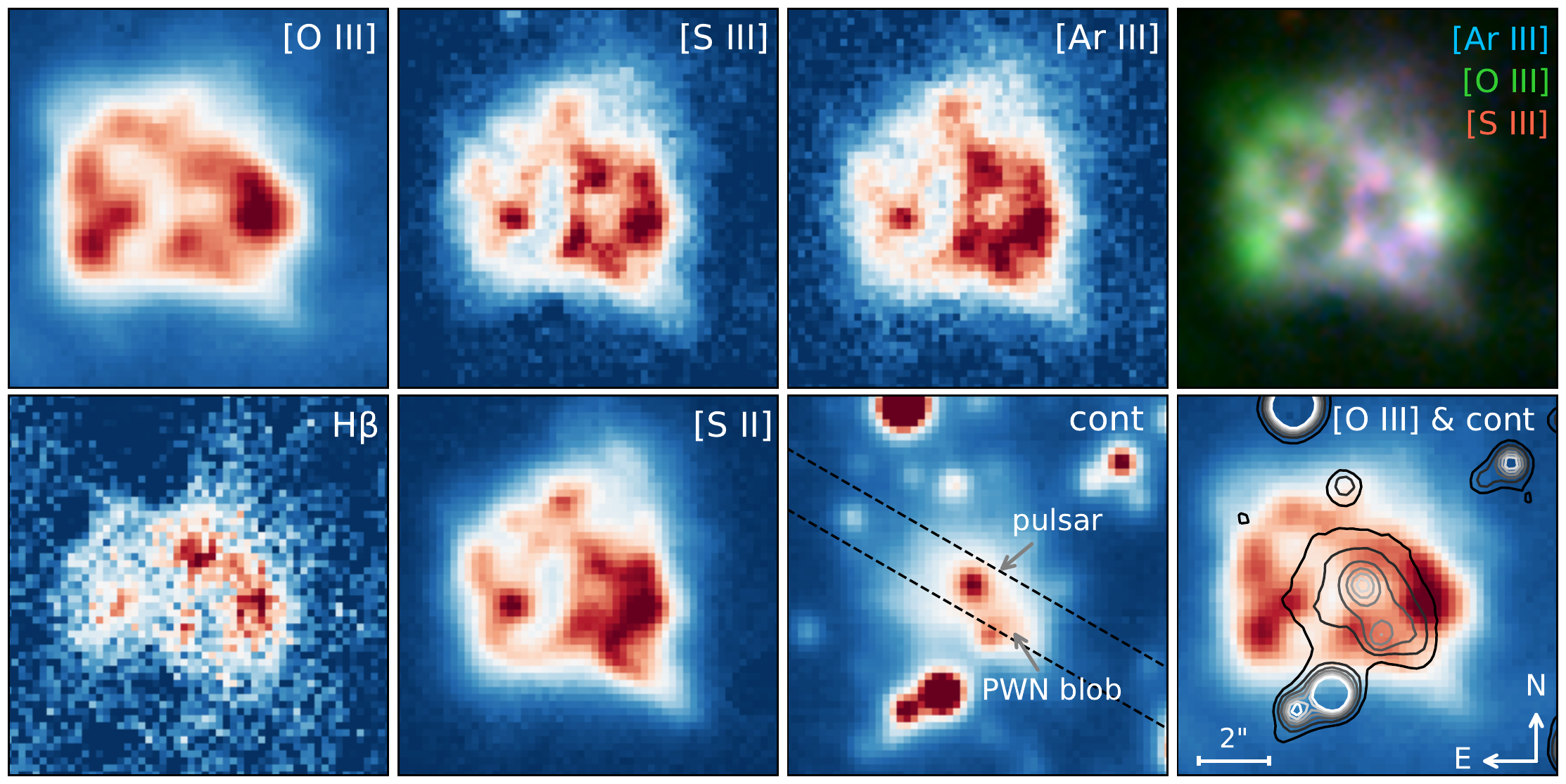}
\caption{MUSE images of the central $11'' \times 11''$ region of SNR~0540. The FOV covers ejecta velocities up to $\sim 1200$~\kms\ from the pulsar in the plane of the sky. Images of emission lines were produced by summing the cleaned cubes over the full spectral region where significant ejecta emission is detected. The H$\beta$ image in the bottom left panel shows some dark areas outside the ejecta region due to residuals from the removal of stars. The location of stars can be seen in the continuum image (obtained around 5300~\AA) in the third panel, bottom row. The positions of the pulsar and a bright blob in the PWN are indicated by arrows in this image, while the dashed black lines show the 1\farcs{5} wide X-shooter slit. The bright spot located just below the lower dashed line near the PWN blob is due to two blended stars. The continuum is also shown as contours superposed on the [\ion{O}{3}] image in the bottom right panel. 
\label{fig:muse_images_inner}}
\end{figure*}

The  ejecta of SNR~0540 are composed of several complex emission components. Below we first give a general overview of the main observed features in the innermost region ($\lesssim 4^{\prime\prime}$) in Section~\ref{sec:results-inner} and on more extended scales ($4-17^{\prime\prime}$) in Section~\ref{sec:results-extended}. A more detailed analysis of the morphology, as well as an age estimate, are presented in Section~\ref{sec:results-analysis}. 

\subsection{Inner region}
\label{sec:results-inner}

\begin{figure*}[t]
\plotone{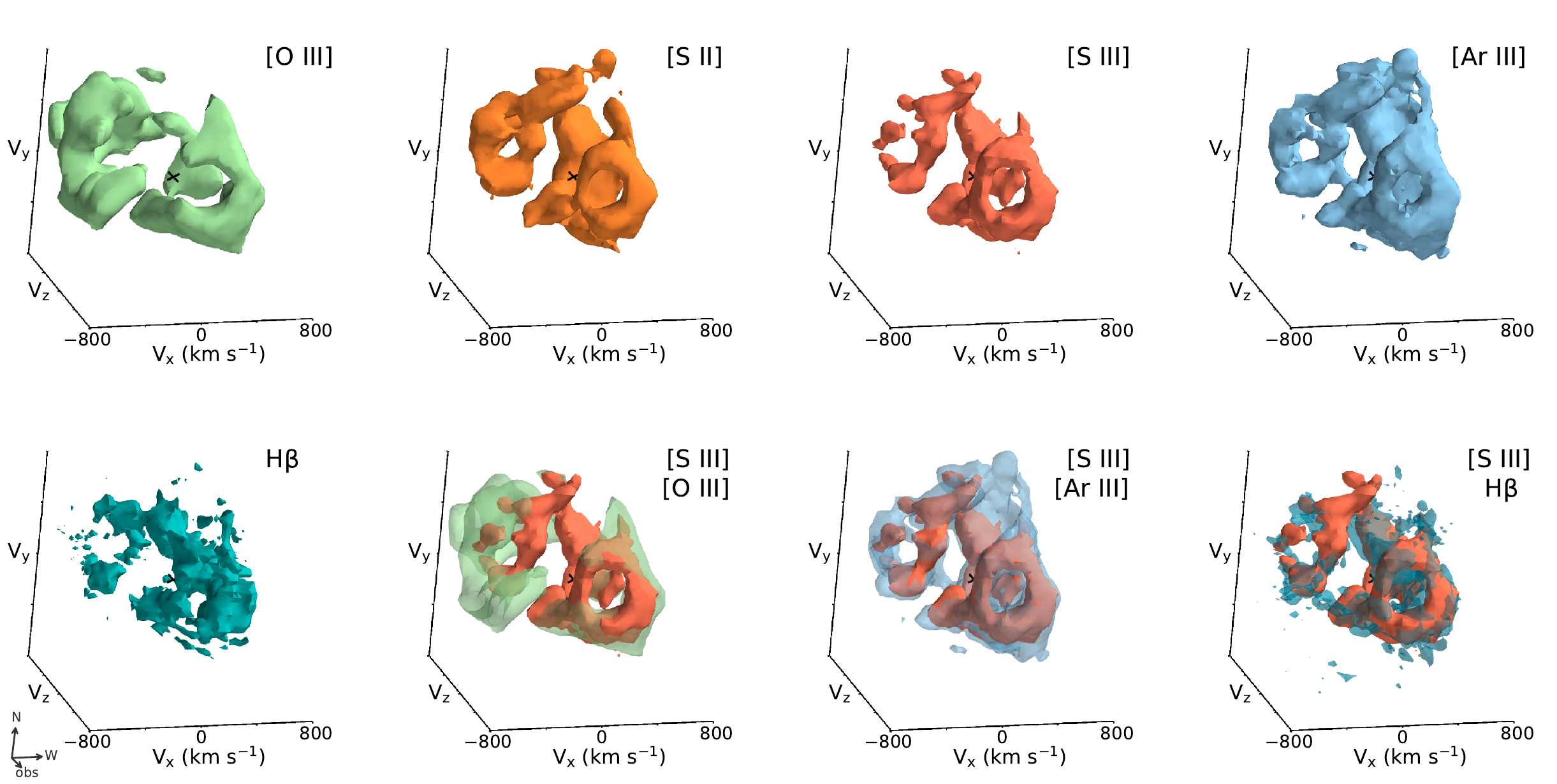}
\caption{ 3D iso-surfaces from MUSE data for the same emission lines as shown in Figure~\ref{fig:muse_images_inner}. The surfaces correspond to $40\%$ of the peak intensity for each line. The last three panels show pairs of emission lines plotted together (using the same color scheme), with one of the iso-surfaces set to be semi-transparent to aid the comparison.  $V_{\rm x}$ and $V_{\rm y}$ are the velocities in the plane of the sky, while $V_{\rm z}$ is the velocity along the line of sight.  The black cross shows the assumed center of explosion. An animated (rotating) version of the third panel in the bottom row is available. The video shows one rotation. 
\label{fig:3dcont_inner}}
\end{figure*}
\begin{figure*}[t]
\plotone{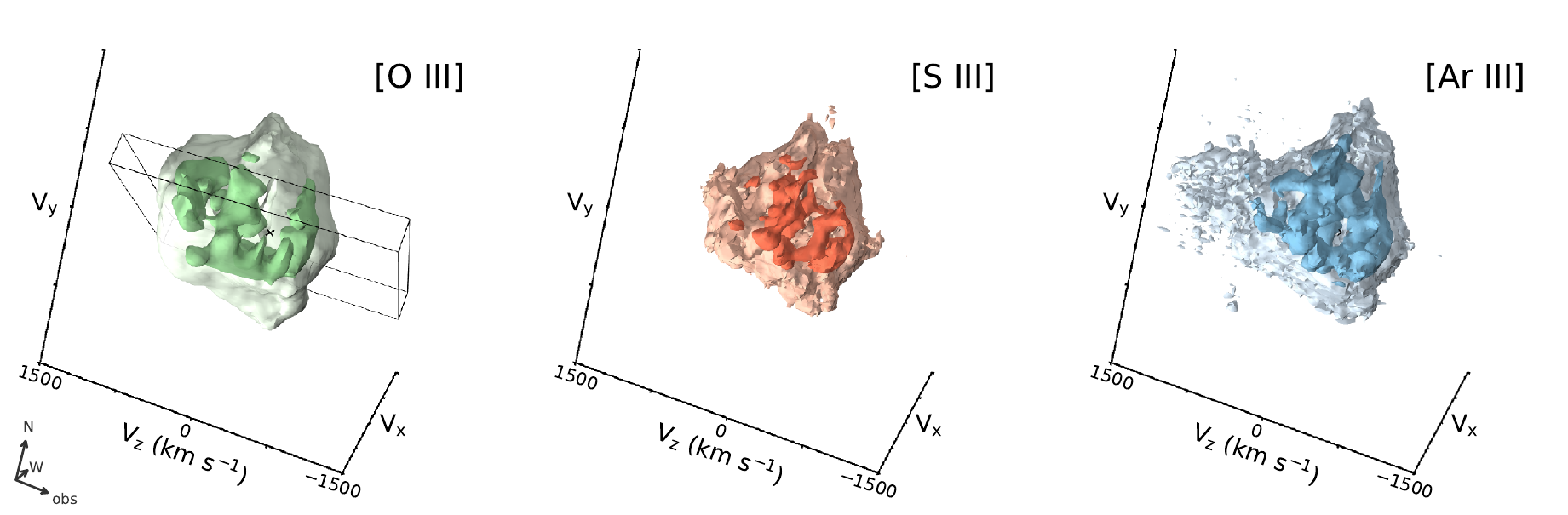}
\caption{3D iso-surfaces from MUSE data highlighting weaker emission in the central region for [\ion{O}{3}], [\ion{S}{3}] and [\ion{Ar}{3}]. The inner surfaces correspond to $40\%$ of the peak intensity, i.e.\ same as in Figure~\ref{fig:3dcont_inner}, while the semi-transparent outer surfaces represent $10\%$ of max. The black box in the left panel illustrates the region covered by the X-shooter slit. The coordinate axes are defined as in Figure~\ref{fig:3dcont_inner}, but 
the axes extend to higher velocities and the viewing angle is different. An animated (rotating) version of the third panel is available. The video shows one rotation. 
\label{fig:3dcont_faint}}
\end{figure*}

\begin{figure*}[t]
\plotone{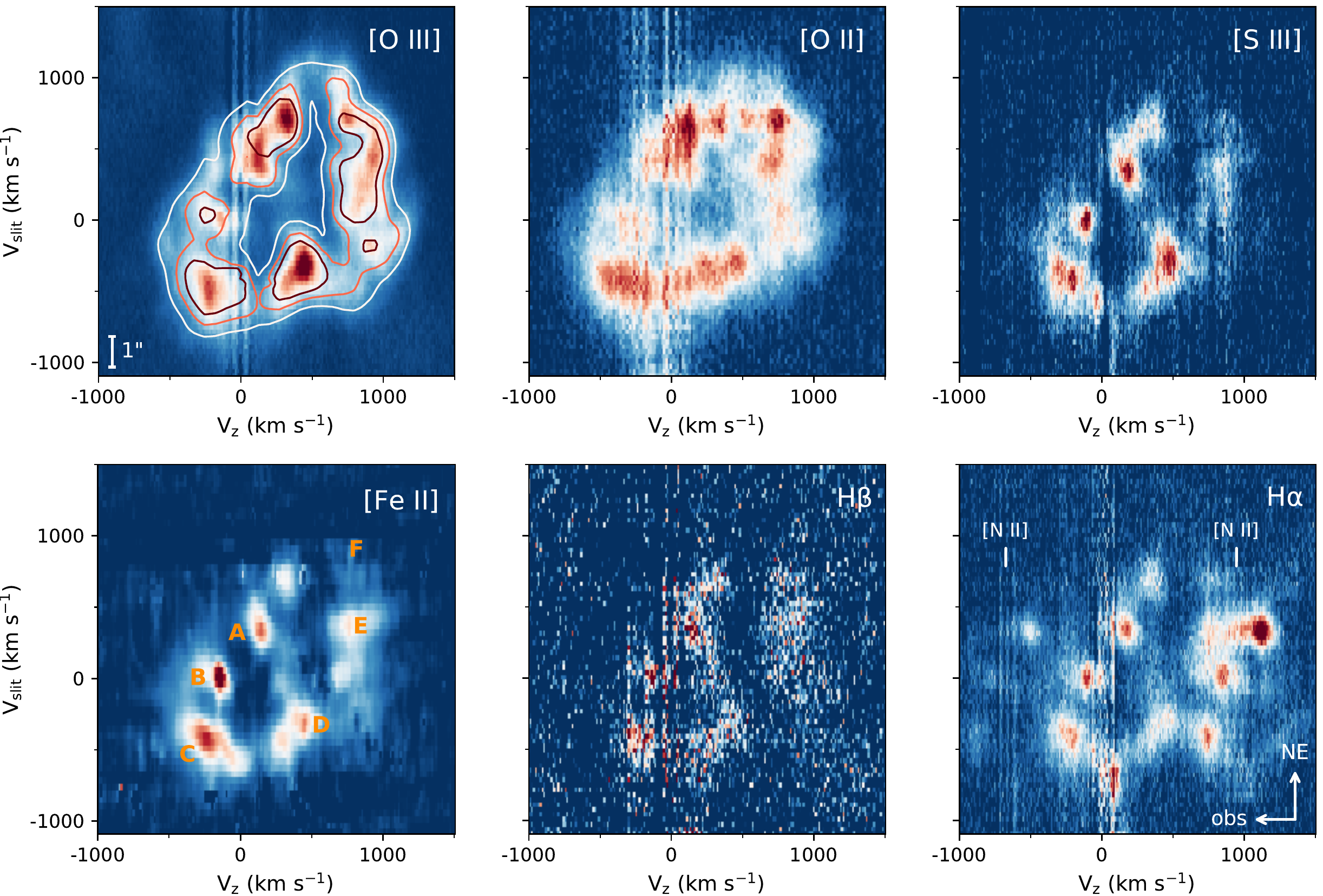}
\caption{2D spectra from X-shooter, showing the region $[-5\farcs{1},+5\farcs{6}]$ around the pulsar position. $V_{\rm slit}$ is the velocity from the center of explosion along the slit, while $V_{\rm z}$ is the velocity along the line of sight. The slit orientation (Figures~\ref{fig:muse_images_inner} and \ref{fig:3dcont_faint}) means that $V_{\rm slit}$ is related to $V_{\rm x}$ and $V_{\rm y}$ in the 3D maps as $V_{\rm slit}  = -(2/\sqrt{3})V_x = 2V_y$. The [O III] contours in the upper left panel represent MUSE data extracted from the region covered by the slit, illustrating the good agreement between the two data sets. The [\ion{O}{2}] $\lambda\lambda 3726, 3729$ doublet in the next panel is centered on the 2nd component. The emission regions labeled A-F in the lower left panel are used for the measurements in Section~\ref{sec:results-clumps}. The images have been stretched in the spatial direction (along $V_{\rm slit}$) in order to give a uniform velocity scale for the assumed age of the remnant.}
\label{fig:xshooter}
\end{figure*}

The strongest line emission from the ejecta is concentrated to a small region within a radius $\sim 4^{\prime\prime}$ ($850$~\kms) from the pulsar. This is illustrated in Figure~\ref{fig:muse_images_inner}, which shows 2D images of the innermost region obtained by summing the cleaned MUSE data cubes for all emission lines. The morphology of the ejecta is dominated by clumpy ring structures, which are similar in all lines, though there are some differences on a more detailed level. The most notable difference is that the [\ion{O}{3}] is slightly more extended in the eastern and western directions. A continuum image is also included in Figure~\ref{fig:muse_images_inner} for comparison. This shows that the line and continuum emission from the PWN overlap in projection, but that the brightest regions are slightly offset (see comparison between [\ion{O}{3}] and continuum in the lower right panel, see also \citealt{Lundqvist2011}). The properties of the continuum emission will be presented in a separate paper (L.~Tenhu et al.,\  in preparation).

We present 3D plots of the inner emission in Figure~\ref{fig:3dcont_inner}. The plots show iso-surfaces corresponding to $40\%$ of the peak intensity for each line, which means that only the brightest structures are seen. The 3D view reveals the ring morphology more clearly than the images and further highlights the similarity between the different lines. The 3D morphology of fainter emission in this region is illustrated in  Figure~\ref{fig:3dcont_faint}. We show only the lines with the best data quality ([\ion{O}{3}], [\ion{S}{3}] and [\ion{Ar}{3}]), omitting H$\beta$  due low S/N and [\ion{S}{2}] due to uncertainties from deblending of the doublet. The plots in Figure~\ref{fig:3dcont_faint} show semi-transparent iso-surfaces corresponding to 10$\%$ of the peak intensity together with the iso-surfaces from Figure~\ref{fig:3dcont_inner}. The faint emission includes a red tail of [\ion{Ar}{3}] that extends to  $\sim 2000$~\kms\ near the center of the remnant. The other lines do not show this tail, though we note that we cannot draw firm conclusions regarding faint redshifted emission in [\ion{S}{3}] due to a higher noise level at these wavelengths. 

Additional information about the morphology of the inner ejecta is provided by the X-shooter 2D spectra in  Figure~\ref{fig:xshooter}. These spectra represent a slice through the full 3D maps, as illustrated in Figure~\ref{fig:3dcont_faint}.  We verified that MUSE and X-shooter give consistent results by extracting a 2D spectrum of [\ion{O}{3}] from MUSE along the X-shooter slit (Figure~\ref{fig:xshooter}, left panel).  The X-shooter velocity maps reveal clumpy ring structures, as expected, with a similar morphology in all lines. This includes [\ion{O}{2}] and  [\ion{Fe}{2}], which are not covered by the MUSE observations. The  [\ion{O}{2}] distribution is slightly more smooth due to blending of the two components of the doublet, which are separated by $224$~\kms. For [\ion{Fe}{2}], we show the 12567~\AA\ line due to its good spectral quality, but note that the 16443~\AA\ line has a fully consistent morphology. 

The higher spectral resolution of X-shooter also allows us to unambiguously identify H$\alpha$ emission in the inner region (Figure~\ref{fig:xshooter}, right panel). A contribution from [\ion{N}{2}]~$\lambda \lambda 6548, 6584$ can be seen on either side of the H$\alpha$ line, with the 6584~\AA\  component having a similar strength as H$\alpha$ \citep[see also][]{Morse2006}.  The overall morphology of the [\ion{N}{2}] emission is consistent with that seen in the other lines. As for MUSE, the X-shooter data reveal some differences between the different lines on a more detailed level. The most obvious difference is stronger [\ion{O}{3}] and [\ion{O}{2}] at the highest redshifts, with the O emission region centered around $V_{\rm z},V_{\rm slit} = (950,-150)$~\kms\ being significantly fainter in the other lines.

\subsection{Extended emission}
\label{sec:results-extended}

\begin{figure*}[t]
\plotone{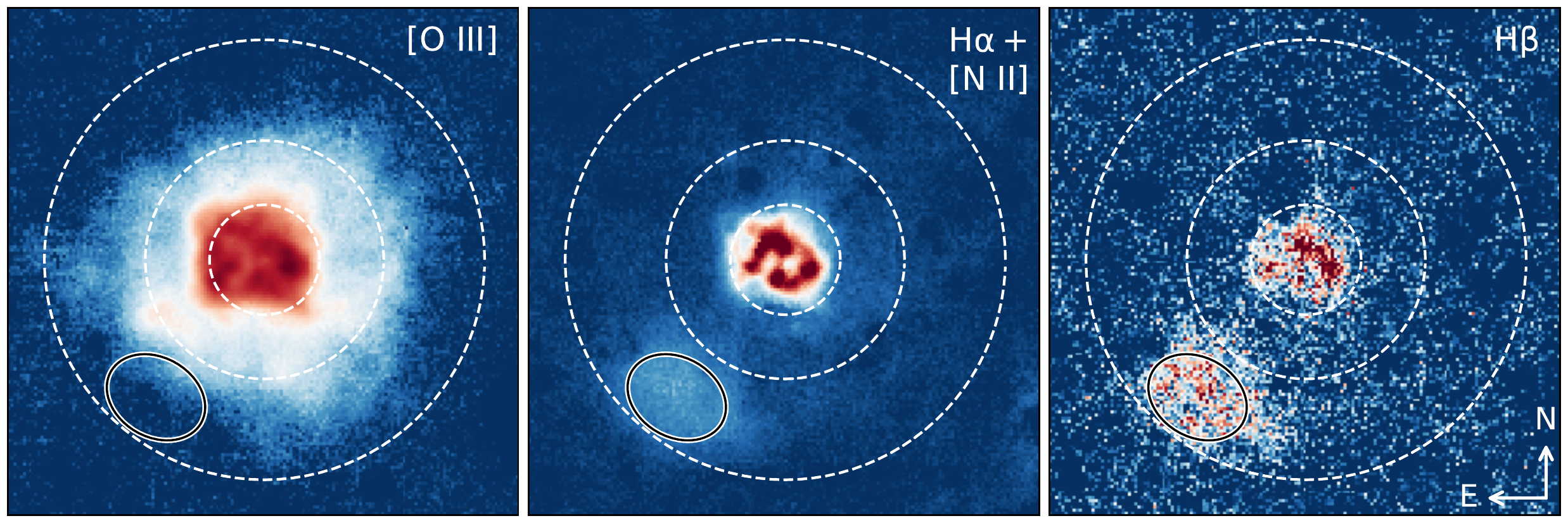}
\caption{MUSE images of extended emission from [\ion{O}{3}],  H$\alpha +$[\ion{N}{2}] and H$\beta$ in SNR~0540. The FOV is $33'' \times 33''$, centered on the position of the pulsar. The images cover Doppler shifts of $\pm 2500$~\kms\ around each line, beyond which no significant ejecta emission is detected. The image in the left panel is scaled by an asinh function to highlight the faint, extended [O III] emission, while the other two images have linear color maps.  The H$\alpha +$[\ion{N}{2}] and H$\beta$ images show a bright emission region in the southeast in addition to the clumpy ring structures in the central region. All other features in these two images are due to residuals from imperfect subtraction of stars and narrow ISM lines. Three circles are plotted for reference to indicate the scale. They have radii of 3.6, 7.8 and $14.4^{\prime \prime}$, which corresponds to $\sim 800$, 1700 and $3100\ \rm{km\ s^{-1}}$, respectively, for an age of 1100~years. An elliptical aperture used for extracting spectra in the region of the H-blob is shown by the black/white line. 
\label{fig:muse_images_big}}
\end{figure*}

We identify spatially extended emission associated with SNR~0540 in [\ion{O}{3}], H$\alpha$ and H$\beta$. This is illustrated by the images in Figure~\ref{fig:muse_images_big}, which were obtained by summing the cleaned MUSE data cubes of the three lines over the central $33'' \times 33''$ region. The FOV corresponds to maximal velocities of $3500$~\kms\ in the plane of the sky, while we only include Doppler shifts within $\pm 2500$~\kms. No significant emission is detected at higher or lower Doppler shifts, though we note that faint, highly red- or blueshifted emission is more challenging to detect if blended with other lines. The cube containing H$\alpha$ has a strong contribution from [\ion{N}{2}] in the inner region, but this is not present in the extended emission, as explained further below. 

\begin{figure*}[t]
\plotone{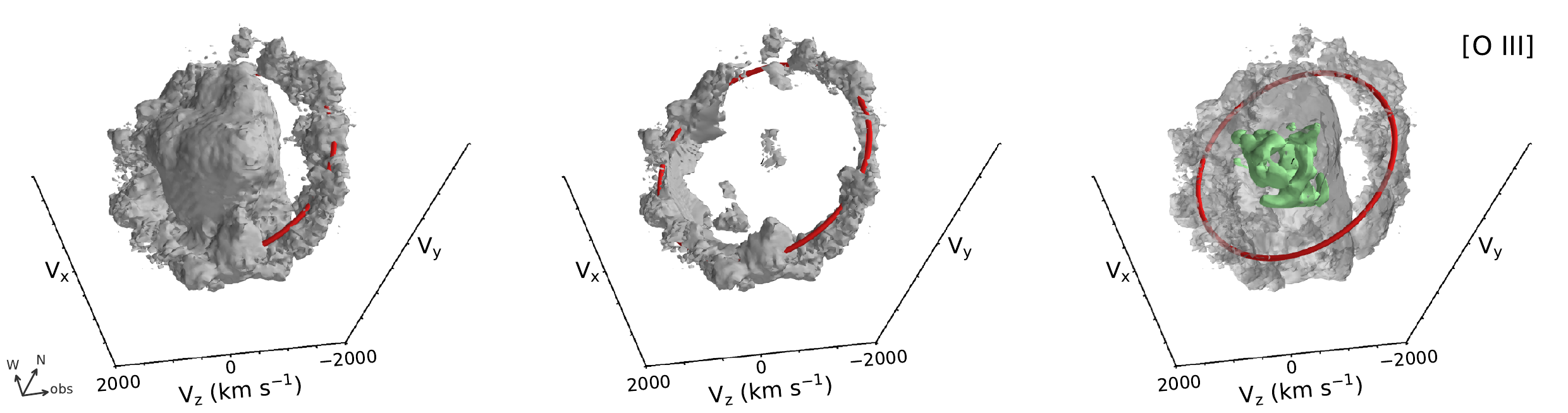}
\caption{3D iso-surfaces from MUSE data showing the faint extended [\ion{O}{3}] emission. Left: the gray surface corresponds to $3\%$ of the peak intensity. Middle: same, but the central region with a radius of $1300$~\kms\ has been removed to highlight the outer structure. Right: the iso-surface at $40\%$ of max (green, cf.~Figure~\ref{fig:3dcont_inner}) is included. The gray surface is the same as in the leftmost panel, but semi-transparent.  A red circle is included for reference to guide the eye (not fitted to data). It has a radius of $1600$~\kms\ and  inclination 40\dg. The coordinate axes are defined as in Figure~\ref{fig:3dcont_inner}. An animated (rotating) version of this figure is available. The video shows one rotation for each panel.   
\label{fig:3dcont_o3_outer}}
\end{figure*}

In the case of [\ion{O}{3}], the emission forms a faint halo that extends to $\sim 10-14^{\prime\prime}$ in different directions. There is some uncertainty in the maximal extent of the weakest emission due to the complex ISM (discussed further in Section~\ref{sec:results-clumps}), but these observations clearly allow us to trace the emission beyond the previously reported $\sim 8^{\prime\prime}$ radius (e.g., \citealt{Morse2006}). With MUSE we can also investigate the 3D morphology of this emission, which is shown in Figure~\ref{fig:3dcont_o3_outer}. The iso-surface in this plot represents only $3\%$ of the peak intensity of the line.   We find that this emission forms a ring-like structure surrounding the inner ejecta. The ring extends outwards from $\sim 1000$~\kms\ and is inclined by $\sim 40^{\circ}$ with respect to the line of sight. The geometry of this structure is quantified further in Section~\ref{sec:results-analysis}, where we also use it to estimate the kinematic age of the remnant. 

\begin{figure}[t]
\plotone{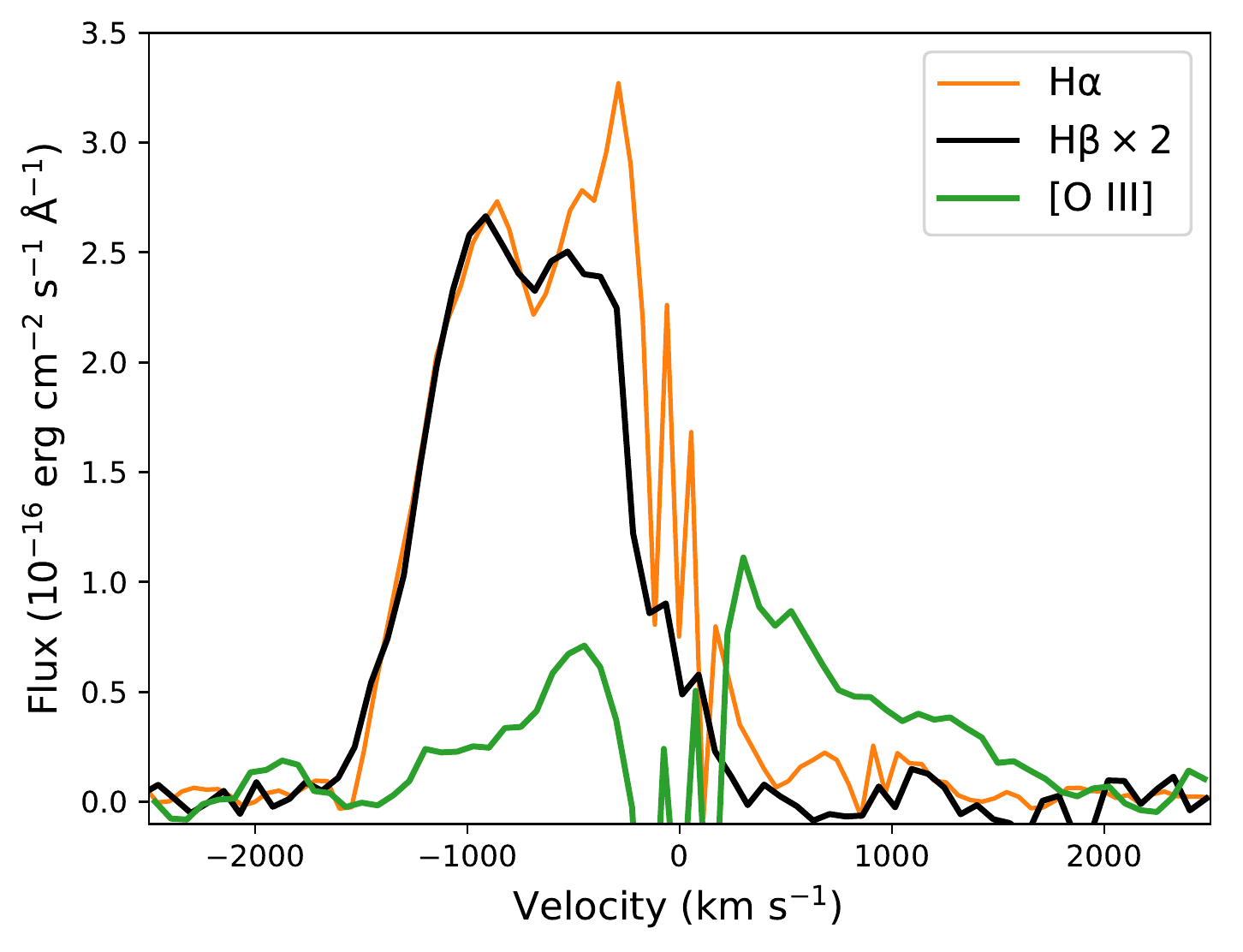}
\caption{MUSE spectra extracted from the elliptical aperture shown in  Figure~\ref{fig:muse_images_big}, covering the H$\alpha$+H$\beta$ emission region in the southeast. The spectrum of [\ion{O}{3}] from the same region is also shown for comparison. The removal of narrow lines from the ISM introduces uncertainties at the lowest velocities, clearly seen in the H$\alpha$ and [\ion{O}{3}] profiles.
\label{fig:blobspectra}}
\end{figure}

\begin{figure}[t]
\plotone{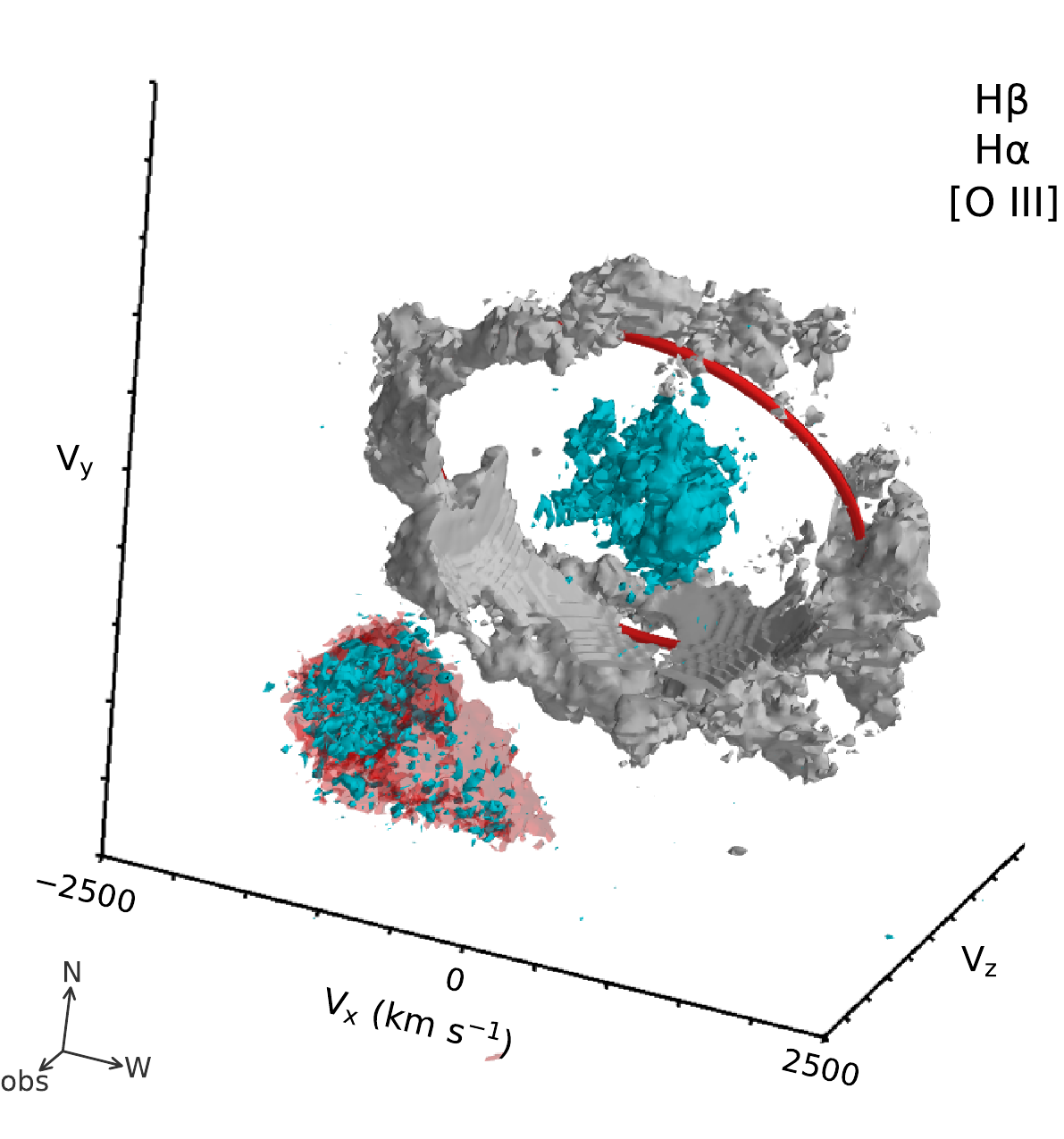}
\caption{3D iso-surfaces from MUSE data showing the high-velocity blob of H and its relation to the extended [\ion{O}{3}] emission. The turquoise surface corresponds to $30\%$ of the peak H$\beta$ emission. The semi-transparent  red surface shows the H$\alpha$ emission in the southeastern part of the remnant. The gray surface is the same as in the middle panel of Figure~\ref{fig:3dcont_o3_outer}, showing the [\ion{O}{3}] emission outside $1300$~\kms. The red ring is the same as in Figure~\ref{fig:3dcont_o3_outer}. The coordinate axes are defined as in Figure~\ref{fig:3dcont_inner}.  An animated (rotating) version of this figure is available. The video shows one rotation. 
\label{fig:3dcont_hblob}}
\end{figure}

The extended emission in the Balmer lines is markedly different from the [\ion{O}{3}] emission. It is characterized by a large blob extended between $\sim 8-14^{\prime\prime}$ southeast of the center (Figure~\ref{fig:muse_images_big}). A spectrum extracted from this region is shown in Figure~\ref{fig:blobspectra}, revealing a broad hump of blueshifted emission extending to $-1600$~\kms. The H$\alpha$ and H$\beta$ profiles show excellent agreement, which implies that there is no significant [\ion{N}{2}] emission blended with H$\alpha$ in this region. Figure~\ref{fig:blobspectra} also shows the spectrum of [\ion{O}{3}] extracted from the same region for comparison. This line is clearly different, showing more emission at redshifted velocities. The relation between the extended [\ion{O}{3}] and H emission is further illustrated by the 3D plot in Figure~\ref{fig:3dcont_hblob}, which shows that the H-blob is located in front of the [\ion{O}{3}] emission with respect to the observer. The [\ion{O}{3}] emission in Figure~\ref{fig:blobspectra} is part of the tilted ring structure in Figure~\ref{fig:3dcont_hblob}. 

\subsection{Analysis of ejecta geometry and age estimate}
\label{sec:results-analysis}

In this Section we analyze the distribution of ejecta emission in more detail. We use a clump-finding algorithm to identify clumps in the 3D cubes (Section~\ref{sec:results-clumpfind}), which we use to estimate the age of the remnant (Section~\ref{sec:results-age}) and to quantify the morphology (Section~\ref{sec:results-clumps}). In Section ~\ref{sec:results-clumps}, we also present full velocity distributions for the different emission lines in MUSE data and provide measures of clump separations and sizes for X-shooter data. All analysis on MUSE data is performed using the original sampling of the cubes, i.e.~without the interpolation used for visualizing 3D iso-surfaces above. \\\\

\subsubsection{Identification of clumps}
 \label{sec:results-clumpfind}
 
We use the FellWalker algorithm \citep{Berry2015} to identify clumps, as implemented in the Starlink CUPID\footnote{\url{https://starlink.eao.hawaii.edu/starlink/CUPID}} package \citep{Berry2007}, which we run using a Python wrapper.\footnote{\url{https://starlink-pywrapper.readthedocs.io/en/latest/index.html}} The algorithm works by performing ``walks" through the data cube, following the path with the steepest gradient until a significant peak is reached. All pixels along paths that reach the same peak are assigned to the same clump. In this way, each pixel that is above a given constant threshold gets assigned to a unique clump. Advantages of this algorithm are that clumps can have any shape and that the results are relatively insensitive to the exact values of the input parameters \citep{Berry2015}. On the other hand, the method does not account for possible overlap between clumps or hierarchical structures. 

We run the clump-finding on the cleaned MUSE data cubes of [\ion{S}{2}], [\ion{S}{3}], [\ion{Ar}{3}], H$\beta$ and [\ion{O}{3}]. The algorithm  works in pixel space without any assumptions about how the pixel coordinates translate to ejecta velocity. The values of several of the input parameters are given in units of the root-mean-square (RMS) noise of the cubes, which we measure in background regions for each line. We set the threshold value, below which pixels are not assigned to clumps (the {\it noise} parameter), to $3\times$RMS, while the largest dip between two peaks that are considered to belong to the same clump ({\it mindip}) is set to its default value of $2\times$RMS. The resolution in the cube is approximately 3.5 pixels (in terms of full width at half maximum, FWHM) in both the spatial and spectral directions. In view of this, we set the minimum number of pixels that a clump can contain ({\it minpix}) to 43 pixels and the {\it maxjump} parameter to 3 pixels.  The latter means that a cube with sides of 6 pixels is searched for higher pixel values when a local maximum is reached, preventing noise spikes on small scales from being interpreted as clumps. All other input parameters are set to their default values. 

We compared the resulting clump assignment with 3D plots of the emission in each line, using both 3D scatter plots and iso-surfaces like those shown in Figure~\ref{fig:3dcont_inner}, ensuring that no noise spikes or other spurious features were assigned as clumps. Following this inspection, we manually rejected some clumps clearly associated with residuals from the ISM subtraction in the case of [\ion{O}{3}], as well as some clumps at high velocities in [\ion{S}{2}] that we associate with residuals from deblending of the doublet. All remaining clumps were analyzed using the {\it findclumps} command of the CUPID package, which provides information about the positions, sizes and fluxes of the clumps.

\subsubsection{Age estimate}
\label{sec:results-age}

\begin{figure}[t]
\plotone{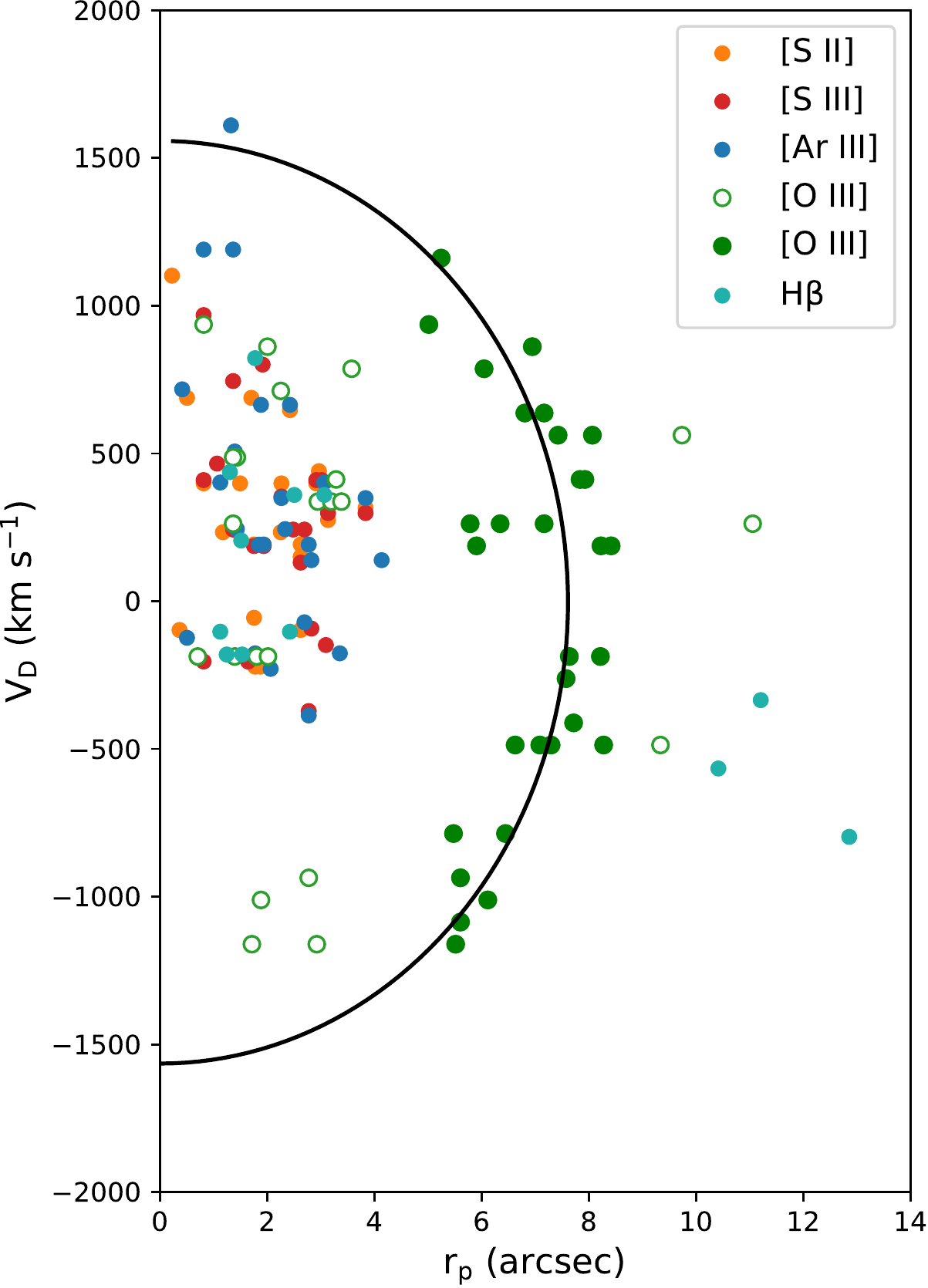}
\caption{Distribution of $v_{\rm D}$ and $r_{\rm p}$ for the peaks of the clumps identified in MUSE data by the FellWalker algorithm. The black line is the best-fit semi-circle used for the age estimate. It was fitted to the [\ion{O}{3}] clumps shown as filled green circles. The [\ion{O}{3}] clumps not included in the fit are shown with open symbols. 
\label{fig:o3clumpage}}
\end{figure}

We use the clumps identified above to estimate a kinematic age for the remnant.  As a first step, we translate the pixel coordinates of the peak position of each clump to Doppler velocities ($v_{\rm D}$) and projected distances ($r_{\rm p}$) from the assumed center (see Section~\ref{sec:methods}). Fitting a semi-circle to the distribution of these observables for knots of ejecta offers a way to estimate the age, as previously done for Cas~A (e.g, \citealt{DeLaney2010}) and N132D \citep{Law2020}. The method assumes freely expanding, spherically symmetric ejecta. Inspection of our clump positions in Figure~\ref{fig:o3clumpage} shows that the inner region (inside $r_{\rm p}<4^{\prime\prime}$) is clearly asymmetric and thus ill-suited for such an estimate. These inner clumps are also likely to have been affected by acceleration by the pulsar wind. On the other hand, the majority of the clumps in the extended [\ion{O}{3}] emission approximately follow a semi-circle, as also expected from the ring-like structure seen in Figure~\ref{fig:3dcont_o3_outer}.  We select the [\ion{O}{3}] clumps located between $4^{\prime\prime}<r_{\rm p}<9^{\prime\prime}$ (filled green circles in Figure~\ref{fig:o3clumpage}) and fit them using the following expression (adapted from \citealt{DeLaney2010}):

\begin{equation}
\left( \frac{r_{\rm p}}{s} \right)^2 + \left( v_{\rm D} - v_{\rm c}   \right)^2 = \left( v_{\rm max} - v_{\rm c}  \right)^2,  
\end{equation}
where the fit parameters are the central Doppler shift of the semi-circle ($v_{\rm c}$),  the maximal velocity when $r_{\rm p} = 0$ ($v_{\rm max}$) and the scale factor ($s$), which translates the $r_{\rm p}$ to a velocity. 

The resulting fit is shown as a black line Figure~\ref{fig:o3clumpage}. The best-fit values are $s = 4.87 \pm 0.50 \times 10^{-3}\ {^{\prime\prime}}/ (\rm{km\ s^{-1}})$, $v_{\rm c} = -4 \pm 63$\,\kms\  and $v_{\rm max} =  1558 \pm 158$\,\kms, with the latter two giving a radius of $1562 \pm 170$~\kms. The scale factor can be translated to an age by combining it with the fact that $1^{\prime\prime}$ corresponds to $7.42 \times 10^{12}$~km at the distance of the LMC (49.6\,kpc, \citealt{Pietrzynski2019}). This gives an age of $1146 \pm 116$~years. For comparison, performing the fit using the positions of the clump centroids instead of their peaks gives an age of $1122 \pm 140$\,years and a radius corresponding to $1646 \pm 231$\,\kms. 

We have also performed the fits after varying the input parameters for the clump-finding algorithm (in particular the {\it maxjump} and {\it mindip} parameters, which affect the number of clumps identified). This results in best-fit ages that are within the statistical uncertainties quoted above. The main observational uncertainty affecting the age estimate is the ISM subtraction, which prevents us from accurately defining clumps with $v_{\rm D} \sim 0$\,\kms. There is also a systematic uncertainty in this age estimate arising from potential deceleration due to interaction or acceleration by the PWN \cite[see e.g.,][for a relevant discussion]{Banovetz2021}. 

Considering these uncertainties, we take 1100 years as the age for SNR~0540. This is in line with previous kinematic age estimates, which are in the range $ 800-1200$~years \citep{Kirshner1989,Morse2006}. Those estimates were obtained by comparing the extent of the emission in images with the Doppler shifts in spectra. The spin-down age of the pulsar is somewhat higher at $\sim 1660$~years \citep{Seward1984}.

\subsubsection{Properties of clumps and rings}
\label{sec:results-clumps}

\begin{figure*}[t]
\plotone{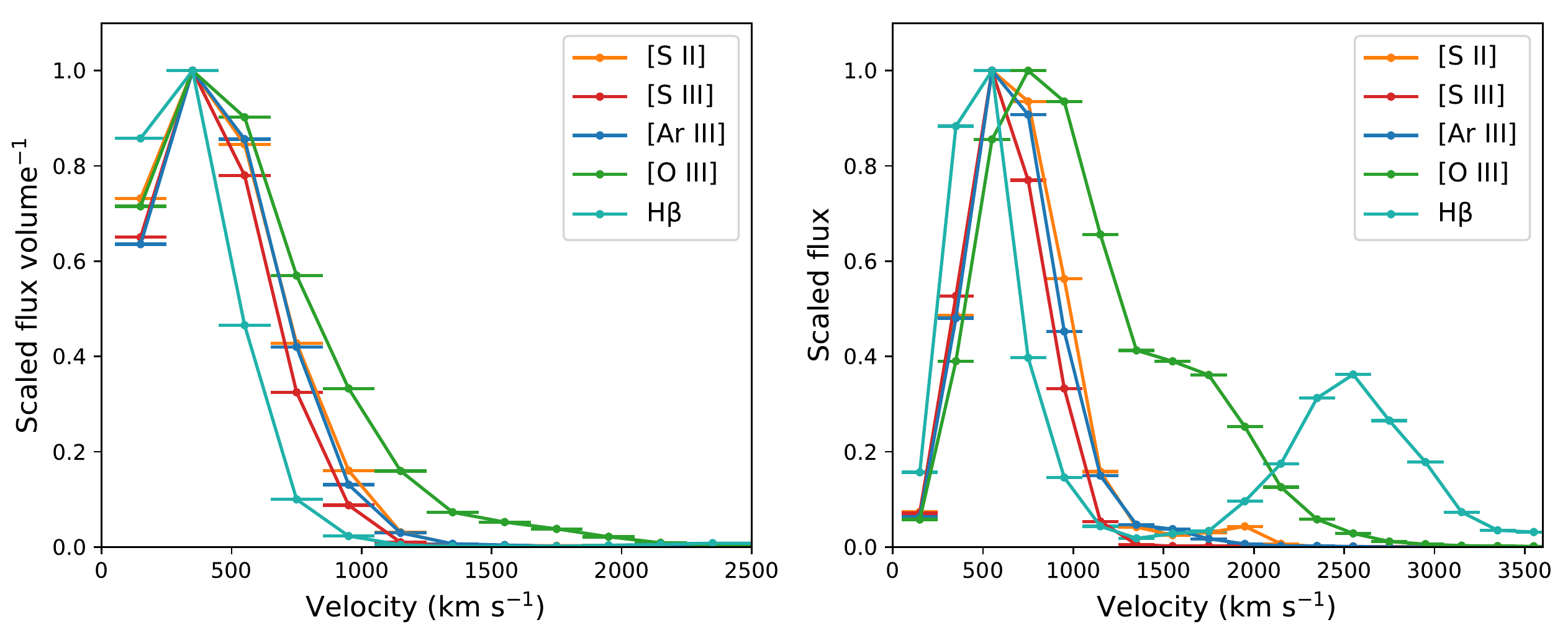}
\caption{Measures of ejecta brightness as a function of velocity for different emission lines. The fluxes in each line were summed in spherical shells of $200\ \rm{km\ s^{-1}}$ width, shown as horizontal error bars. Left: Profiles showing the flux divided by the volume of each shell. Right: Corresponding profiles without dividing by the volume. This gives more weight to the outer shells, highlighting the velocity distribution of the extended emission. Note the different extent on the x-axis in the two panels. The analysis was performed on MUSE data, which is also the case for all other figures presented in this Section (Figures~\ref{fig:veldist_segm} - \ref{fig:o3clumps}). 
\label{fig:veldist}}
\end{figure*}
\begin{figure*}[t]
\plotone{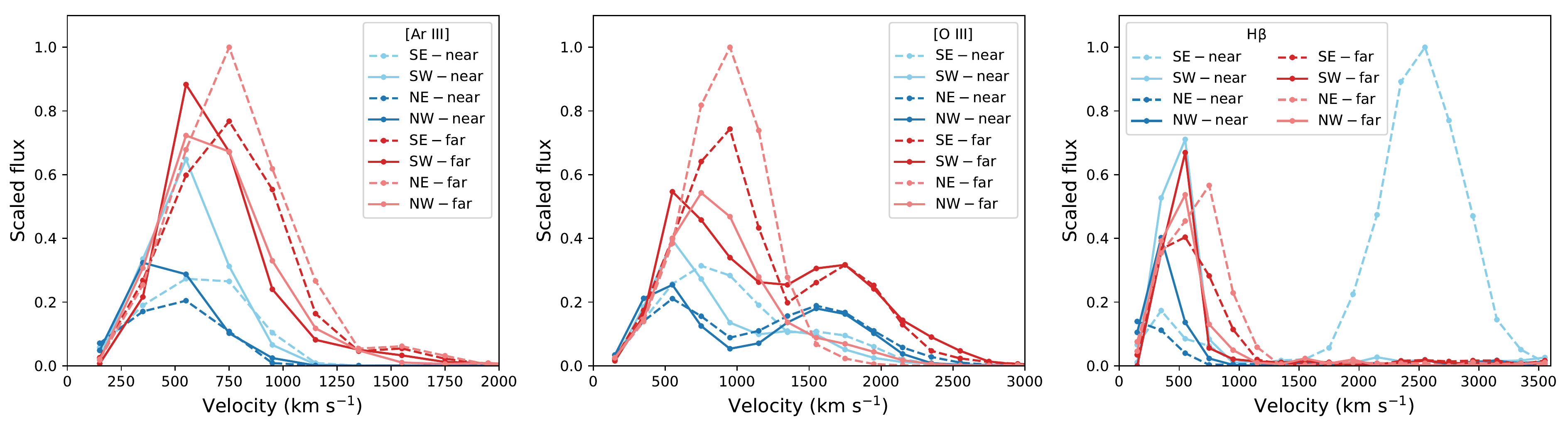}
\caption{Flux profiles in different directions for [\ion{Ar}{3}] (left), [\ion{O}{3}] (middle) and H$\beta$ (right). The fluxes were summed in shells of $200\ \rm{km\ s^{-1}}$ width in the directions indicated by the labels, with ``near" and ``far" indicating blueshifted and redshifted emission, respectively. The shells are the same ones indicated by the horizontal error bars in Figure~\ref{fig:veldist}. The sum of all directions are shown in the right panel of Figure~\ref{fig:veldist}. Note the different extent on the x-axis in the three panels.
\label{fig:veldist_segm}}
\end{figure*}

\begin{figure*}[t]
\plotone{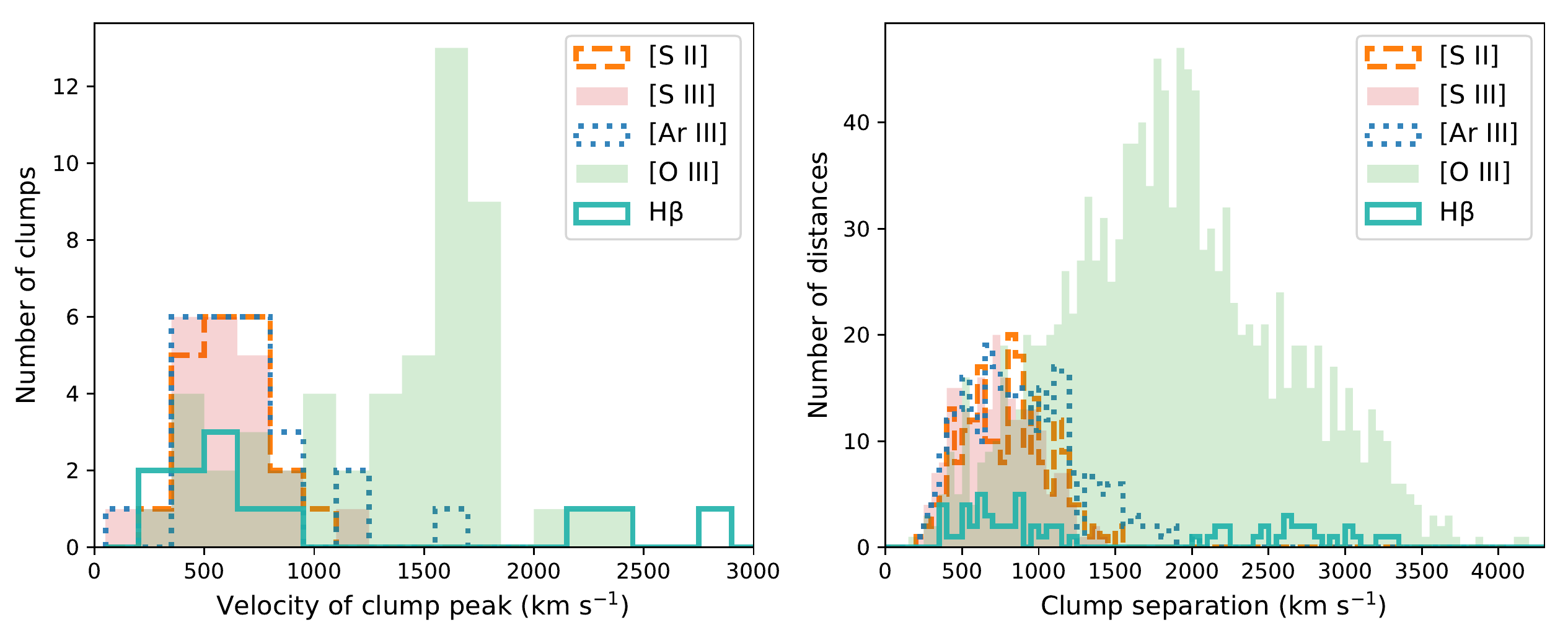}
\caption{Distributions of clumps identified for different emission lines. Left: velocities of the peak positions of the clumps. Right: separations in velocity between all pairs of clumps in the left panel.  
\label{fig:clump_pos}}
\end{figure*}

\begin{figure}[t]
\plotone{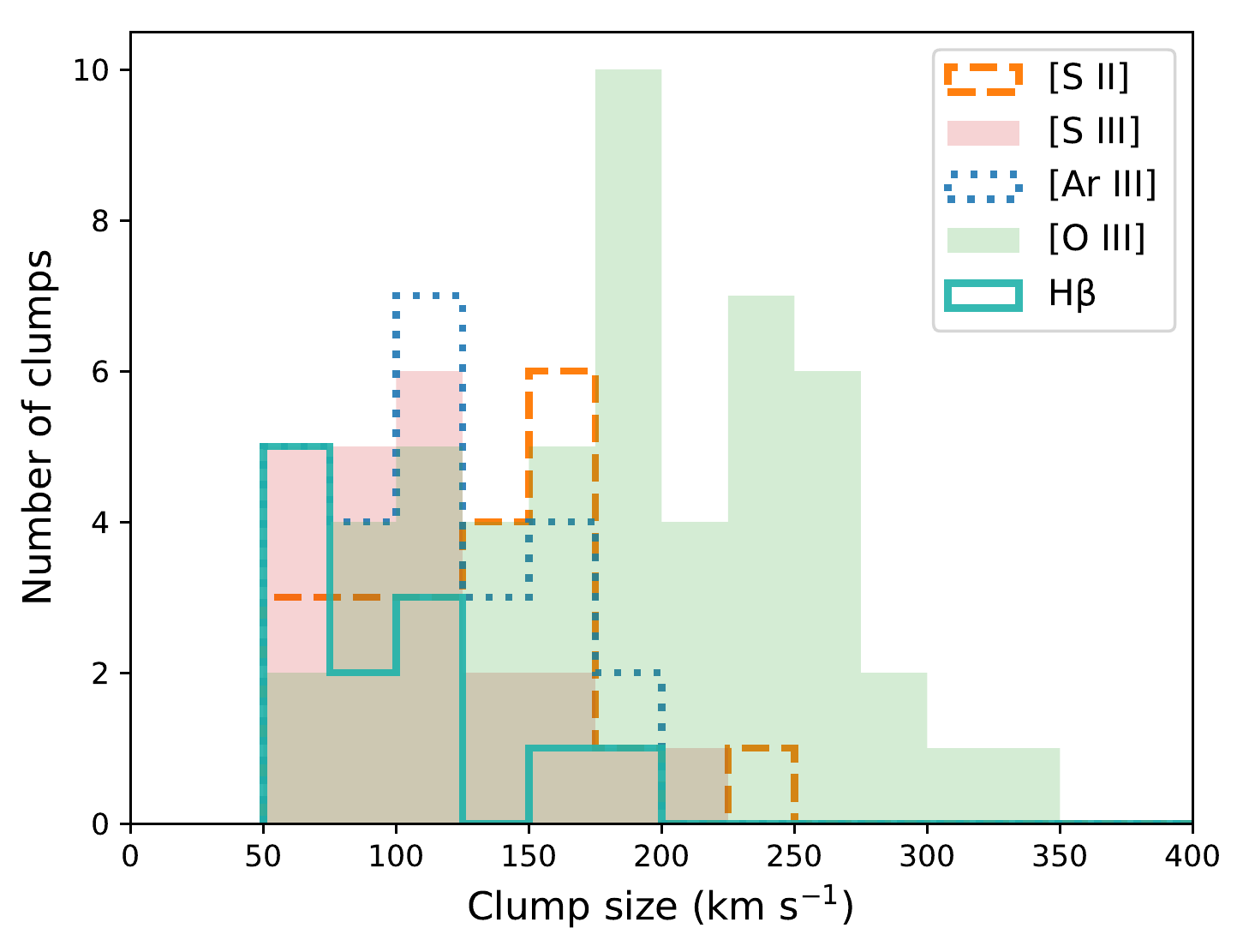}
\caption{Distribution of clump sizes for all emission lines. The plotted sizes are the averages of the RMS sizes in the three dimensions ($V_{\rm x},\ V_{\rm y}$ and $V_{\rm z}$), deconvolved with the resolution (see text for details). The smallest clump sizes defined in this way are in the range $56-68$~\kms\ for the different lines, but the lower boundary of the first bin in the histogram is set to 50~\kms\ for all lines for visual clarity.  
\label{fig:clump_size}}
\end{figure}

\begin{figure*}[t]
\plotone{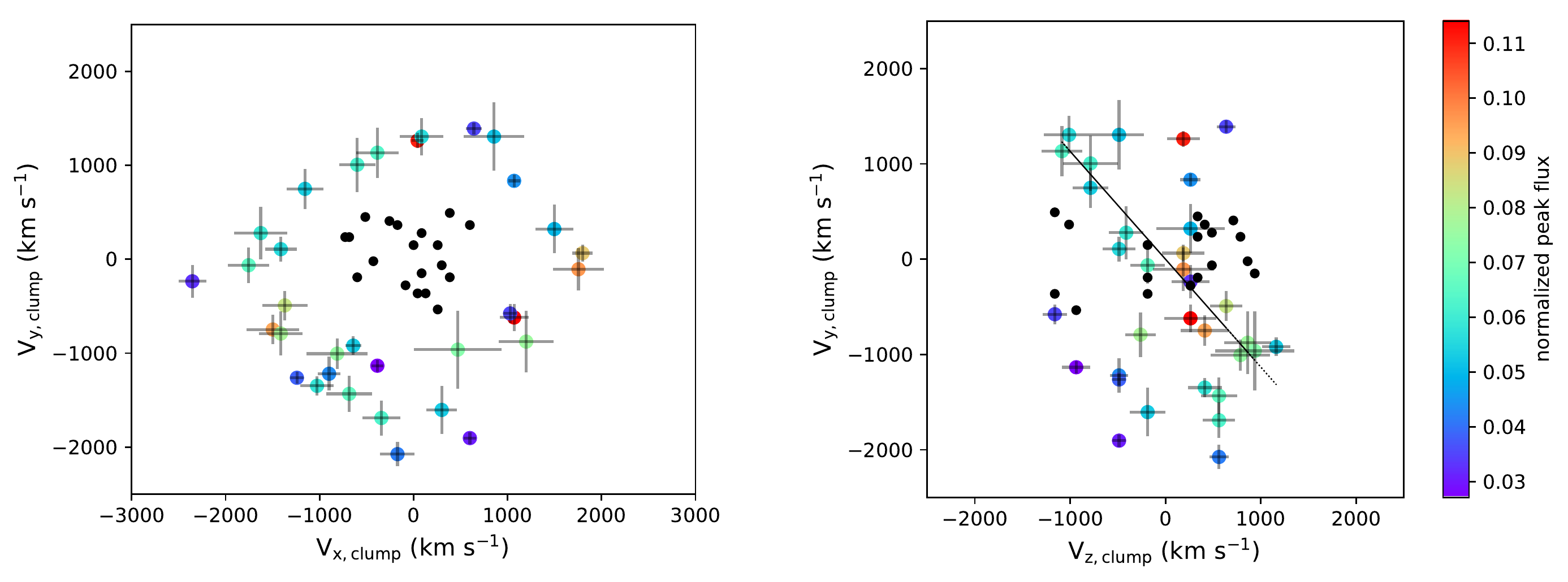}
\caption{Positions of clumps identified in the [\ion{O}{3}] emission projected in the plane of the sky ($V_{\rm x} - V_{\rm y}$, left) and as seen from the ``side" ($V_{\rm y} - V_{\rm z}$, right). The clumps shown in black are all located in the inner $4^{\prime\prime}$, while the clumps located further out are color-coded by their peak fluxes (normalized such that the brightest pixel, belonging to one of the black points in the central part, has a value of 1). The error bars represent the RMS sizes of the clumps in the different directions (omitted for the black points in the innermost region for visual clarity). The black line in the right panel was fitted to the colored clumps with normalized fluxes $>0.05$. It corresponds to an inclination of $\sim 40$\dg\  with respect to the line of sight. 
\label{fig:o3clumps}}
\end{figure*}

The characteristic velocities of the main components of the ejecta are summarized in Table~\ref{tab:ejectasum}, with further details provided in the text below.  All velocities quoted are 3D velocities measured from the assumed center of explosion for an age of 1100 years, unless otherwise stated. 

To quantify the full velocity distribution of the observed emission, we sum the flux in the MUSE data cubes in spherical shells of $200$~\kms\  width, starting at $50$~\kms. The width of the bins was selected considering the resolution, which corresponds to FWHM~$\sim 150$~\kms\ in the spatial direction and ranges from FWHM~$\sim 90$ at [\ion{S}{3}] to $\sim 170$~\kms\ at H$\beta$.  We only include pixels with flux values $>3\times$RMS above the background. The left panel of Figure~\ref{fig:veldist} shows the flux in each shell divided by its volume. The profiles are completely dominated by the inner part, with all lines peaking in the velocity bin that covers $250-450$~\kms. The profiles of [\ion{Ar}{3}], [\ion{S}{2}] and [\ion{S}{3}] fall off similarly with increasing velocity, reaching $\sim 1200$~\kms, while [\ion{O}{3}] shows more emission at higher velocities, with a tail extending to just over $2000$~\kms. The H$\beta$ profile is narrower than the other lines, with a higher flux per volume in the innermost bin. This is likely influenced by the lower S/N and lower spectral resolution for this line.  The high-velocity blob of H$\beta$ is not noticeable in this profile because it occupies a small part of the total volume of the outer shells. 

The H-blob is instead clearly seen in the right panel of Figure~\ref{fig:veldist}, where the flux in each shell has not been normalized by volume. This shows that the emission in the blob extends from $\sim 1500-3500$~\kms. The same plot also shows  [\ion{O}{3}]  extending to $\sim 3000$~\kms\ and a faint tail of [\ion{Ar}{3}] reaching $\sim 2000$~\kms\ (the small bump in [\ion{S}{2}] at this velocity is affected by residuals from deblending of the doublet). 

Figure~\ref{fig:veldist_segm} shows the flux profiles of [\ion{Ar}{3}], [\ion{O}{3}] and H$\beta$ divided into eight directions, which provides information about the asymmetries. In the inner regions, all lines show a general trend of higher fluxes and higher velocities on the redshifted side; the magnitude of the peaks on the red- and blueshifted sides differ by up to a factor of 5 and the positions of the peaks differ by up to 400~\kms. The two S lines (not shown) exhibit the same trends.  Outside the inner region, the [\ion{O}{3}] profiles show clear bumps at high velocities (peaking at $\sim 1600-1700$~\kms) in the north-blueshifted and south-redshifted directions, which are associated with the extended ring-like structure seen in Figure~\ref{fig:3dcont_o3_outer}. The flux at these velocities is a factor 7 higher in the south than in the north on the redshifted side, while it is a factor 2 higher in the north than in the south on the blueshifted side, which gives a measure of how pronounced the ring is. Finally, the H$\beta$ profiles show a bright peak corresponding to the blueshifted blob in the southeast, but there is no corresponding emission in any other direction. 

It is interesting to compare these flux profiles with the distribution of clumps identified by the clump-finding algorithm. A similar number of clumps (21--25) were obtained for [\ion{S}{2}], [\ion{S}{3}] and [\ion{Ar}{3}], a total of 52 clumps were obtained for [\ion{O}{3}] due to the added contribution from the extended emission, while only 12 clumps were identified in H$\beta$ due to the lower S/N. The left panel of Figure~\ref{fig:clump_pos} shows the distributions of velocities for the clump positions (defined as the location of the peak flux of each clump). All emission lines show a concentration of clump positions in the range $300-800$~\kms, which is comparable to the main peak of the flux profiles. In the case of [\ion{O}{3}], there is a prominent peak in the clump distribution at 1600~\kms, corresponding to the extended ring, as well as a tail of clumps reaching a maximal velocity of 2400~\kms. For H$\beta$, the clump-finding algorithm identifies three clumps with comparable peak fluxes in the high-velocity H blob, with centers at 2300, 2440 and 2860~\kms, respectively (see Table~\ref{tab:ejectasum} for details). 

The right panel of Figure~\ref{fig:clump_pos}  shows the distribution of velocity separations between all pairs of clumps for each line. The distributions are broad, as expected from the fact that we see structures on several different scales in the 3D plots. To estimate the typical size of the small rings in the inner region, we determine the distance between clumps located on opposite sides of the blueshifted ring in the southwest of the ejecta (this is the most well-defined ring, clearly seen in Figure~\ref{fig:3dcont_inner}). We also measure the distance between clumps on opposite sides of the rings in X-shooter data. In this case the clump positions are simply taken as the peaks in the 2D spectra shown in Figure~\ref{fig:xshooter}. The results of these measurements are summarized in Table~\ref{tab:ejectasum}, showing that the typical diameter of the small rings are $\sim 500-600$~\kms. As an approximate measure of the size of the inner cavity, we identify the peak of the emissivity in different directions for [\ion{Ar}{3}] (as in the left panel of Figure~\ref{fig:veldist_segm}, but normalized by volume). The peaks are in the bins centered at 550~\kms\ in the southwest and northeast on the redshifted side, at 150~\kms\ in the southeast and northeast on the blueshifted side, and at 350~\kms\ in the remaining four directions.

The clumps defined by the FellWalker algorithm have irregular shapes. As an approximate measure of their sizes, we use the RMS size in each direction, as provided by the  {\it findclumps} command in CUPID. This corresponds to the standard deviation for the ideal case of a clump with a Gaussian profile. To be able to directly compare the size estimates in different lines, we subtract the resolution in quadrature from the RMS size in each dimension. The relevant resolutions are 1/2.35 of the FWHM values provided at the beginning of this subsection. In $\sim5\%$ of all clumps, the RMS returned by {\it findclumps} is smaller than the resolution, which we deal with by setting the relevant RMS to the value of the resolution. There are no systematic differences between the sizes in the three dimensions and we therefore show the average values in Figure~\ref{fig:clump_size}. 

The clump sizes measured in this way are $\sim 60-200$~\kms\  for all lines with the exception of [\ion{O}{3}], which shows clump sizes extending up to 330~\kms.  There are no systematic differences between the sizes of [\ion{O}{3}] clumps in the inner and outer regions, as shown in Table~\ref{tab:ejectasum}. We caution that this is a rather crude estimate of the size of the irregular clumps and that the clump sizes are more sensitive to the input parameters of the clump-finding algorithm than the distribution of clump positions. To provide another measure of the clump sizes, we also fit Gaussians to the X-shooter spectra of the clumps, focusing on [\ion{Fe}{2}] and H$\alpha$, which offer the best spectral resolution. These results are summarized in Table~\ref{tab:ejectasum}, showing that the smallest clump is unresolved at $\sim 30$~\kms. 

We investigate the properties of the clumps in the extended [\ion{O}{3}] emission in more detail in Figure~\ref{fig:o3clumps}, which shows the projected positions of the clumps in the plane of the sky and as seen from the side. The sizes and fluxes of the clumps located outside $4^{\prime\prime}$ are also indicated in the figure. This illustrates that the extended emission is more complex than a simple ring or torus, as also evident from several previous plots (e.g., Figures~\ref{fig:3dcont_o3_outer}, \ref{fig:3dcont_hblob} and \ref{fig:o3clumpage}). In particular, there are a number of clumps in the south that fall below the plane of the ring, some of which are on the blueshifted side. It is notable that the clumps located furthest from the ring have low fluxes and small sizes. This also agrees with Figure~\ref{fig:veldist_segm} (middle panel), which shows peaks in the flux profiles at high velocities only in the north-blueshifted and south-redshifted directions, i.e. along the plane of the ring. From inspection of the full 3D maps, it is clear that the clumps located away from the ring are not noise spikes. On the other hand, it is likely that the exact positions of some of these clumps are affected by uncertainties due to the removal of narrow lines from the surrounding medium. 

As an example, the clump located furthest to the east with $V_{\rm x}, V_{\rm y} = (-2350,-240)$~\kms\ (Figure~\ref{fig:o3clumps}) overlaps with a highly complex region of the ISM. The Doppler shift of its brightest point is only $V_{\rm z} =260$~\kms, which is likely affected by contamination from the ISM. However, the clump extends to redshifts of $\sim 1000$~\kms, with a peak at $\sim 500$~\kms\ in its integrated line profile, showing a clear association with SNR~0540. To obtain a conservative measure of the highest-velocity [\ion{O}{3}] associated with the remnant, we investigate the velocity distribution of all pixels assigned to clumps after excluding positions with $|V_{\rm z}|<500$~\kms. We find that this distribution extends to $3000$~\kms, consistent with the flux profiles in Figures~\ref{fig:veldist} and \ref{fig:veldist_segm}.     

To obtain an approximate estimate of the inclination of the main ring, we fit a straight line to the $V_{\rm z} - V_{\rm y}$ positions of all the clumps that have peak fluxes above a threshold of 0.05 in normalized units. The exact value of the flux cutoff is somewhat arbitrary, but motivated by the fact that most of the faintest clumps do not follow the main ring structure. The best-fit line has a slope of $-1.1\pm0.2$  (Figure~\ref{fig:o3clumps}, right panel), which corresponds to an inclination of 42\dg. This estimate is clearly dependent on which clumps are included in the fit, but we note that an inclination of 40\dg\  gives a good visual agreement  with the 3D contours seen in Figure~\ref{fig:3dcont_o3_outer} (red circle in that figure). 

\begin{deluxetable*}{lll}[t]
\tablecaption{Summary of ejecta components \label{tab:ejectasum}}
\tablecolumns{3}
\tablenum{1}
\tablewidth{0pt}
\tablehead{
\colhead{Component} & 
\colhead{Velocity (\kms)}  &
\colhead{Notes} 
}
\startdata
{\bf Inner PWN (all lines)}	&	& \\
	velocity range &	$\lesssim 1000$	&	Figures~\ref{fig:veldist} and \ref{fig:clump_pos}  \\
	inner cavity &	$150-550$ (median 350)	&	Distance from center to max flux volume$^{-1}$ in different directions	 \\
	ring diameters&	$460\pm 20$	&	Distances between clumps on opposite sides of southwest ring in MUSE \\ 
	 			& $650\pm 10$ 	&	Distance A-E in X-shooter (Figure~\ref{fig:xshooter}) \\ 
				& $620\pm 10$ 	&	Distance B-D in X-shooter (Figure~\ref{fig:xshooter}) \\ 
				& $1630\pm 10$ 	&	Distance C-F in X-shooter (Figure~\ref{fig:xshooter}) \\ 
	clump sizes\tablenotemark{a}  &	 $60-200$ (median 120) 	&	Clumps in MUSE,  all lines except [\ion{O}{3}]\\  
	&	 $90-330 $ (median 200)  &	Clumps in MUSE for [\ion{O}{3}] inside $4^{\prime\prime}$ \\  
	&	$30-80$	&	X-shooter, $\sigma$ from fitting  Gaussians to spectra of clumps in [\ion{Fe}{2}] and H$\alpha$ \vspace{1mm} \\ \hline
{\bf Extended [\ion{O}{3}] } &	& \\	
	 velocity range & $1000-3000$	& 	Figures~\ref{fig:veldist}, \ref{fig:veldist_segm} and \ref{fig:clump_pos} \\
	ring radius  & 1600 (inclined by $\sim 40$\dg) &		Figures~\ref{fig:veldist_segm} and \ref{fig:clump_pos} \\
	 clump sizes\tablenotemark{a}	&	 $70-320 $ 	(median 190)&	 Clumps in MUSE for [\ion{O}{3}] outside $4^{\prime\prime}$ \vspace{1mm} \\ \hline
{\bf High-velocity H blob} &	& \\
	 velocity range	 & 1$500-3500$ & Figures~\ref{fig:veldist} and \ref{fig:veldist_segm} \\
	position &	 2300 [-1500,  -1650,   -570]	&	Positions of clump peaks in $V$ [$V_{\rm x}, V_{\rm y}, V_{\rm z}$]  \\
	  &	 2420  [-940, -2200,   -340]	& \\
	 &	 2860 [-1580,  -2250, -800]	& \\
	 clump sizes\tablenotemark{a} 	& 180, 120, 70	&	For the clumps specified in the row above \\
\enddata
\tablecomments{The listed values are approximate characteristic velocities of the different components.}
\tablenotetext{a} {The quoted sizes for MUSE are the defined as in Figure~\ref{fig:clump_size}, i.e.\ the average of the RMS sizes in the three dimensions. The sizes have been deconvolved by the resolution for both MUSE and X-shooter.} 
\end{deluxetable*}

\section{Discussion}
\label{sec:disc}

The 3D distribution of ejecta in SNe carry information about the progenitor star and the explosion mechanism, as shown by numerical simulations \cite[e.g.,][]{Gabler2021,Orlando2021}. Reconstructed 3D maps of ejecta emission therefore offer a powerful way of evaluating theoretical models, though the observations will of course also be affected by the relevant energy sources and resulting ionization structures.  Below we discuss these effects and the interpretation of the observations of the innermost region (Section \ref{sec:disc-inner}), the extended [\ion{O}{3}] emission (Section~\ref{sec:disc-oring}) and the H-blob (Section~\ref{sec:disc-hblob}),  before finally placing SNR~0540 in the context of other SNRs in Section~\ref{sec:disc-comp}. 

\subsection{The innermost region}
\label{sec:disc-inner}

The emission in the inner region of SNR~0540 is powered by the PWN. In particular, the ejecta located inside $\sim 4^{\prime\prime}$ are expected to be shocked by the pulsar wind \cite[e.g.,][]{Williams2008}. This is supported by our detection of [\ion{Fe}{2}], which is a shock-tracer.  We find typical velocities of $\sim 500$~\kms\ in this region, assuming freely expanding ejecta. \cite{Williams2008} estimate a velocity of 20~\kms\ for the shocks in the dense clumps, which is small compared to the free-expansion speeds. The observations show that the morphology in this region is broadly similar in all lines, including different ionization states of the same element. This indicates that the rings and clumps likely reflect the true density distribution of ejecta, rather than variations in temperature and ionization level. 

On a more detailed level, there are some differences between the different emission lines. The most notable differences in the MUSE observations are between [\ion{O}{3}] compared to [\ion{S}{2}], [\ion{S}{3}] and [\ion{Ar}{3}]. We find that the rings extend to higher velocities in [\ion{O}{3}]  (by up to $\sim 200$~\kms\ in some directions, Figures~\ref{fig:muse_images_inner}, \ref{fig:3dcont_inner}, \ref{fig:veldist}), that [\ion{O}{3}] has larger maximal clump sizes (Figure~\ref{fig:clump_size}), and also lacks the faint red tail seen in [\ion{Ar}{3}] (Figure~\ref{fig:3dcont_faint}, where the S lines are uncertain). These differences may reflect the origin in different nuclear burning zones of the progenitor, with S and Ar being located further in (as part of the the Si/S layer) than O.  While these layers have been mixed in the explosion, the  small significant differences between the lines indicate that some of the original structure has been retained. Similar conclusions have previously been reached for Cas~A, where infrared observation of the ejecta interior to the reverse shock show differences between O and Si in some locations \citep{Isensee2010}. 

Cas A also shows Fe-rich ejecta that extend beyond S-rich ejecta \citep{Hughes2000,Milisavljevic2013}, which simulations explain as fingers of Fe ejecta pushing through the lower-density Si/S layer \citep{Orlando2021}. Our X-shooter observations show no indication of such inversion, with [\ion{Fe}{2}] and [\ion{S}{3}] having very similar distributions, which are slightly different from [\ion{O}{3}] on the redshifted side (Figure~\ref{fig:xshooter}). A likely reason for this difference between SNR~0540 and Cas A is that the inversion of layers in Cas~A was enhanced by the passage of the reverse shock \citep{Orlando2021}.      

The mixing of different elements observed in the inner region of SNR~0540 is not solely due to the explosion, but is also expected to have been enhanced as a result of low-velocity (and low-density) ejecta being swept up by the PWN shock. However, this process clearly only operates in the outward direction, which leads us to conclude that elements that originate further out in the progenitor, including H, must have been mixed down to velocities of only a few 100~\kms\ in the explosion itself. Similarly low velocities for H have been observed also in SN~1987A \citep{Kozma1998,Larsson2019} and reproduced in simulations of neutrino-driven explosions \citep{Utrobin2019}. 

Simulations also predict the presence of so-called Ni bubbles, where the radioactive decay of $^{56}$Ni pushes the ejecta outward, compressing them into a denser shell \citep{Li1993,Blondin2001,Gabler2021,Orlando2021}. This process may help explain the observed morphology of the inner ejecta in SNR~0540, as bubbles intersected by the pulsar shock would manifest as rings in the observations. In this interpretation, the observed ring diameters of $\sim 500-600$~\kms\ (Table~\ref{tab:ejectasum}) provide a measure of the size of Ni bubbles in the innermost region. The shell of a Ni bubble is expected to be clumpy \citep{Basko1994} and this clumpy structure will be further enhanced by fragmentation resulting from the interaction with the shock \citep{Williams2008}, which agrees with our observation of clumpy rings. The Ni bubbles are expected to be filled with Fe (the decay product of Ni), so the fact that we observe [\ion{Fe}{2}] emission in the clumpy rings rather than in the interior of the rings (Figure~\ref{fig:xshooter}), is consistent with a picture where the shock has swept up the material inside the bubbles. 

In addition to the small-scale asymmetries in the form of rings and clumps, the inner ejecta of SNR~0540 show a global asymmetry in the sense that the redshifted ejecta have higher fluxes and higher velocities than the blueshifted side. This has been noted in previous observations \citep{Kirshner1989,Serafimovich2005,Morse2006,Sandin2013} and is likely caused by asymmetries in the explosion and/or the energy input from the pulsar. Net motion of the progenitor relative to the local ISM may also play a role in explaining this asymmetry, as originally discussed by \cite{Kirshner1989}. However,  this is unlikely to be the dominant effect, as the offset is up to $\sim 400$~\kms\  (Figure~\ref{fig:veldist_segm}), implying that the progenitor would have been unbound from the LMC.

\subsection{The extended [\ion{O}{3}] emission}
\label{sec:disc-oring}

The MUSE observation has allowed us to characterize the [\ion{O}{3}] emission that surrounds the inner nebula in unprecedented detail. The faintest emission can be traced to velocities of $\sim 3000$~\kms, but we find that the emission is dominated by an irregular ring-like structure with a radius of $\sim 1600$~\kms\ and inclination $\sim 40$\dg. Using clumps in this ring, we obtain an estimate of the age of the remnant of 1100 years. This agrees reasonably well with the spin-down age of $\sim 1660$~years for the pulsar \citep{Seward1984}, providing evidence that the [\ion{O}{3}] emission most likely originates from freely expanding SN ejecta rather than pre-explosion mass loss. Furthermore, the [\ion{O}{3}]-emitting material is likely to be photoionized by the synchrotron emission and radiative shocks of the PWN \citep{Williams2008}, implying that the observed velocities have not been altered by shock interaction.  

If the [\ion{O}{3}] ring were part of the CSM, the observed properties would point to an eruptive mass-loss event with high velocity shortly before the explosion. Eruptions from luminous blue variable stars can reach velocities $> 1000$~\kms\ in some cases \cite[e.g.,][]{Smith2008,Smith2011}, but the typical progenitors are very massive ($>20~\rm M_\odot$), which is unlikely to be the case for SNR~0540. Shells of CSM have also been observed in Type Ib/c SNe with lower-mass progenitors, such as SN~2001em \citep{Chugai2006}, SN~2014C \citep{Milisavljevic2015_14C} and SN~2019oys \citep{Sollerman2020}. However, these kinds of shells are H-rich, whereas the ring in SNR~0540 emits only in [\ion{O}{3}]. Another possible complication with interpreting the [\ion{O}{3}]  ring as CSM is that it would have had to survive the passage of the main blast wave many years ago. 

In our preferred interpretation that the [\ion{O}{3}] emitting material is SN ejecta, the observed velocities are expected to reflect the location of the O zone of the progenitor. In line with this, the simulations of neutrino-driven SN explosions by \cite{Wongwathanarat2015} shows a shell of O at comparable velocities for the 15~\msun red supergiant model ``W15" (their figure 14). The reason we observe a ring-like structure rather than a shell of O in SNR~0540 may be due to localized energy input from the pulsar/PWN, a CSM geometry that constrains the expanding ejecta into a plane, asymmetries in the explosion and/or rotation of the progenitor. In the scenario where the ring reflects the density distribution rather than a localized source of ionization, we can obtain a simple estimate of the density enhancement in the ring plane from the flux profiles in Figure~\ref{fig:veldist_segm}. We assume that the temperature does not vary with position and consider the fact that the emissivity scales as density squared. A comparison of the red- and blueshifted fluxes in the northern and southern hemispheres then shows that the density enhancement along the ring is 1.4 and 2.6 in the north and south, respectively. 

In terms of understanding the origin of the ring, it is interesting to note that a torus of [\ion{O}{3}] emission at a velocity of $\sim 1700$~\kms\ has been observed in the O-rich remnant N132D \citep{Law2020} and that ALMA observations of SN~1987A have revealed a torus of CO, also with a radius of $\sim 1700$~\kms\ \citep{Abellan2017}. These observations of rings/tori of O at similar velocities in different remnants raises the possibility that this structure is intrinsic to the ejecta distribution, rather than reflecting the details of the energy input (pulsar/PWN in SNR~0540, reverse shock in N132D and radioactive decay of $^{44}$Ti in SN~1987A). An example of such an explanation is that the rings are connected to the rotation of progenitor stars with similar masses. At the same time we note that rings are not ubiquitous. For example, the O-rich SNR G292+1.8 also shows reverse-shocked [\ion{O}{3}]  knots at $\sim 1700$~\kms, but the distribution appears to be consistent with a shell in this case \citep{Ghavamian2005}. 

There are currently no clear predictions for how the ejecta distribution will be affected by rotation, as numerical simulations of explosions evolved to late times have not been performed for rotating progenitors. However, it is interesting to note that a flattened distribution of ejecta resembling Cas~A  has been obtained in 3D simulations of a neutrino-driven explosion of a non-rotating star evolved to the remnant stage \citep{Orlando2021}. In this case, a thick torus-like ejecta distribution reflects a large-scale asymmetry of Fe at the time of shock breakout.

\subsection{The H-blob}
\label{sec:disc-hblob}

The MUSE observation revealed a large blob of H$\alpha$ and H$\beta$ emission located southeast of the pulsar, at velocities in the range $1500-3500$~\kms. The velocities are comparable to those observed for [\ion{O}{3}], but the emission regions of the two lines do not overlap, with the H-blob being located on the near side of the [\ion{O}{3}] ring with respect to the observer. In our calculations of the full 3D velocities we have assumed that the blob is freely expanding ejecta, which is supported by the similar distance from the center and Doppler shifts as observed for the extended [\ion{O}{3}]. However, other explanations cannot be ruled out, including that it is CSM or a blown-off envelope of a binary companion. 

There are no other broad lines detected from the blob region that can provide additional information about its properties. Unfortunately, any possible emission from \ion{He}{1} $\lambda 5876$ cannot be analyzed due to the gap in the wavelength coverage of MUSE. To estimate the H mass in the blob, we use the observed luminosity of H$\beta$. While the S/N of H$\beta$ is low, we consider this luminosity more reliable than that of H$\alpha$ due to better removal of narrow lines (Figure~\ref{fig:blobspectra}) and the possibility of a low level of contamination by [N~II] in the H$\alpha$ profile. 

Assuming fully ionized H, the recombination emission from H$\beta$ can be approximated as \citep{Osterbrock1989}
\begin{equation}
j_{\rm H\beta} = 1.24 \times 10^{-25} \left( \frac{T}{10^4}\right)^{-0.91} n_{\rm e} n_{\rm p}\ \rm{erg\ s^{-1}\ cm^{-3}},  
\end{equation}
where $T$ is the temperature and $n_{\rm e}$ and $n_{\rm p}$ are the electron and proton number densities, respectively. For a blob with volume $V$, where the electrons and protons have the same filling factor, $f$, the luminosity can be expressed as
\begin{equation}
L_{\rm blob, H\beta} = j_{\rm H\beta} V f^2  = \frac{j_{\rm H\beta} M_{\rm blob,p} f}{m_{\rm p} n_{\rm p}} \ {\rm erg\ s^{-1}},
\end{equation}
where we have used the fact that the mass of H in the blob is given by $M_{\rm blob,p} = V m_{\rm p} n_{\rm p} f$, with $m_{\rm p}$ being the proton mass. 
Combining these expressions and using the measured H$\beta$ luminosity of $L_{\rm blob, H\beta} = 1.0 \pm 0.1 \times 10^{33}\ \rm{erg\ s^{-1}}$, the mass of the blob can be expressed as 
\begin{equation}
M_{\rm blob,p} = 6.8\ (n_{e} f)^{-1}  \left( \frac{T}{10^4}\right)^{0.91}\ {\rm M}_\odot\ . 
\label{blobmass}
\end{equation}
The values of the variables on the right hand side are unknown, but given that $f<1$, a substantial mass of H is implied. 

In the scenario where the blob originates from the H-envelope of a binary star, one would expect $\lesssim 10\%$ of the total mass of a close companion to have been blown off, depending on the properties of the explosion and companion star \citep{Liu2015,Rimoldi2016,Hirai2018}. As an example, for a typical temperature of  $T=10^4$~K, equation~\ref{blobmass} gives $M_{\rm blob,p}>0.6\ {\rm M}_\odot$ for an effective electron density of $n_{\rm e} f = 10\ {\rm cm^{-3}}$, implying a massive companion. We stress that these numbers are uncertain and that the effective density is unknown, but the example serves to illustrate that the binary interpretation cannot be ruled out. 

Advantages of this interpretation is that it provides natural explanations for the high velocities (with the spectra showing strong emission extending to blueshifts of $-1600$~\kms, Figure~\ref{fig:blobspectra}), as well as the lack of emission in any other direction around the pulsar. By contrast, the high velocities are challenging for any explanation related to CSM (or ISM), as discussed in the context of [\ion{O}{3}] above. In addition, even if there are examples of stars with highly asymmetric mass-loss, such as VY~CMa \citep{Smith2001}, the H-blob would constitute an extreme example of asymmetric CSM. A similar consideration also applies to the interpretation of the blob as ejecta -- a single large blob of high-velocity H clearly stands out as exceptional when compared to asymmetric ejecta observed in other SNRs.  

A possible explanation for the highly localized emission within the ejecta and CSM interpretations is that the blob is energized by a pulsar jet, implying that any H located away from the jet axis would be invisible to us. A jet-axis along the northwest-southeast direction has been proposed based on the X-ray morphology \citep{Gotthelf2000}, which is also supported by changes in the polarization angle along the perpendicular axis \citep{Lundqvist2011}. The blob is indeed located at the southeastern end of the proposed jet direction. However, the morphological X-ray features attributed to a jet are strongest in the northwest, where no H emission is seen. It thus seems unlikely that the observations can be explained solely by localized energy input from a jet, at least not without also invoking a highly asymmetric distribution of ejecta or CSM.  Another difficulty for the jet interpretation is the lack of any apparent X-ray or radio emission from the region of the blob \citep[see e.g.,][]{Brantseg2014} as might be expected from shocks driven by a jet, though we stress that a quantitative analysis would need to be carried out to constrain models.

\subsection{SNR 0540 in the population of SNRs}
\label{sec:disc-comp}

SNR~0540 belongs to the class of O-rich remnants, which are generally thought to originate from stripped-envelope SNe. Our detection of H$\alpha$ and H$\beta$ in the innermost region clearly demonstrates that SNR~0540 was a Type~II explosion. However, the innermost region only contains a small fraction of all the ejecta, which makes it difficult to draw any firm conclusions regarding the total mass of H and hence whether the SN was a Type~IIb or a IIP. The high-velocity H-blob would support the latter if associated with ejecta, but we cannot exclude alternative explanations as discussed above. 

Cas~A, the most well-studied O-rich SNR, is known to originate from a Type~IIb explosion based on light-echo observations \citep{Krause2008}, which also reveal evidence for asymmetries \citep{Rest2011}. Like SNR~0540, it shows rings of ejecta, but on markedly larger scales at higher velocities of $\sim 5000$~\kms\  \citep{Milisavljevic2015}. The lower ejecta velocities in SNR~0540 may reflect differences in the progenitor, with SNR~0540 having a more massive H envelope more in line with expectations for a Type~IIP SN. On the other hand, most of the emission in Cas~A is powered by the reverse shock and there is no PWN that can energize its lowest-velocity ejecta, which will naturally lead to some differences in the range of observed velocities. Assuming that the reverse shock in SNR~0540 is located just inside the radio/X-ray shell, it would be at an ejecta velocity of $\sim 6400$~\kms. To date there is no clear evidence for optical emission associated with a reverse shock in SNR~0540, but we will present a detailed search for this using these MUSE observations in future work. 

Compared to other O-rich remnants, SNR~0540 shows some similarities with N132D and G292, as already mentioned above. The remnant E0102 in the Small Magellanic Cloud is another interesting member of this class \citep{Dopita1981}. It has been reported to show rings of ejecta that resemble the expectations for Ni bubbles \citep{Eriksen2001}. Recent MUSE observations of this remnant also revealed fast-moving knots of H$\alpha$ and H$\beta$ emission with Doppler shifts in the range $-1800 - +800$~\kms\  \citep{Seitenzahl2018}. This shows that E0102 was a Type II explosion (likely a Type~IIb), though we note that the H knots are smaller and distributed over a wider range of directions than the large blob of H that we identify in SNR~0540. 

It is also illuminating to compare the ejecta distribution in SNR~0540 with the Crab. Despite the rather different progenitors, the two remnants are similar in terms of the pulsar and PWN. The Crab was recently mapped in 3D, revealing an overall heart-shaped morphology with filamentary structures in a honeycomb distribution \citep{Martin2021}. While the large-scale morphology appears to be constrained by a pre-existing circumstellar disc, it is also clearly elongated along the axis of the pulsar torus. This is different from SNR~0540, where the direction of the pulsar axis is uncertain and the overall 3D morphology has a more irregular shape. However, there are similarities in terms of the small-scale morphology, which includes ring-like structures that are resolved in much greater detail in the Crab than in SNR~0540. The fact that similar structures are seen also in remnants that lack PWNe, like Cas~A, indicate that they likely reflect intrinsic asymmetries in the explosions.

\section{Conclusions}
\label{sec:conclusions}

We have analyzed the 3D distribution of ejecta in SNR~0540 using observations from the integral-field spectrograph MUSE. We focused on the strongest emission lines, which are  H$\beta$, [\ion{O}{3}]~$\lambda \lambda 4959, 5007$, H$\alpha$, [\ion{S}{2}]~$\lambda \lambda 6717, 6731$,  [\ion{Ar}{3}]~$\lambda 7136$  and [\ion{S}{3}]~$\lambda 9069$. These observations were complemented by 2D spectra from X-shooter, which provides high spectral resolution and also covers the emission lines  [\ion{O}{2}]~$\lambda \lambda 3726, 3729$ and [\ion{Fe}{2}]~$\lambda 12567$. We constructed 3D velocity maps assuming freely expanding ejecta, centered at the position of the pulsar and the average redshift of the local ISM. The clump-finding algorithm FellWalker was used to analyze the 3D maps. From these observations we identify three main emission components, summarized below. 

\begin{itemize}

\item {\bf Inner PWN.}  This region covers the innermost ejecta with velocities $\lesssim 1000$~\kms, thought to have been shocked by the pulsar wind. The ejecta are distributed in clumpy rings that surround an inner cavity. The cavity extends $150-550$~\kms\ in different directions, the rings have diameters $500-600$~\kms\ and the clumps have typical RMS sizes $\sim 100$~\kms, with the smallest clumps reaching the resolution limit of $\lesssim~30$~\kms\ in X-shooter data. The clumps are thus smaller than the voids. There is also a global asymmetry, with higher fluxes and velocities on the redshifted side. All emission lines have broadly similar morphologies, including H$\alpha$ and H$\beta$, providing clear evidence that H was mixed down to low velocities in the explosion.  The mixing is also expected to have been enhanced by the pulsar wind sweeping up low-velocity ejecta. On a more detailed level, there are small differences between different lines. The most significant difference is that O extends slightly further out (by up to $\sim 200$~\kms) than the lines from S, Ar and Fe in some directions, which may reflect the origin in different layers of the star. 

\item {\bf Extended [O~III] emission.} Faint [O~III] emission surrounds the inner PWN in all directions. The emission is dominated by an irregular ring-like structure with a radius of $\sim 1600$~\kms\ and  inclination $\sim 40$\dg, though faint emission can be traced to $\sim 3000$~\kms. Using clumps identified in the [O~III] ring, we estimate a kinematic age of $1146\pm116$~years for the remnant.  We interpret the ring as ejecta rather than pre-SN mass loss based on the agreement between this kinematic age and the pulsar spin-down age ($\sim 1660$~years), together with the fact that the high velocities match the expected location of the O zone of the progenitor. The reason why the O forms a ring rather than a shell may be due to localized energy input from the pulsar/PWN, a CSM geometry that constrains the expanding ejecta into a plane, asymmetries in the explosion and/or rotation of the progenitor. 

\item {\bf High-velocity H-blob.}  A bright blob of H$\alpha$ and H$\beta$ emission is located southeast of the pulsar, extending between radial distances $8-14^{\prime \prime}$ and exhibiting a broad spectral profile with blueshifted velocities reaching $-1600$~\kms. If interpreted as freely expanding ejecta, the full velocities span the range $1500-3500$~\kms. The blob is positioned in front of the [O~III] ring with respect to the observer and there is no corresponding emission in any other direction around the pulsar. We tentatively favor an interpretation as ejecta based on the location and velocities, but note that alternative explanations cannot be ruled out, including the possibility that it is part of a blown-off envelope of a binary companion.    

\end{itemize}

These results highlight the importance of asymmetries and mixing in SN explosions. They also add to the growing evidence that rings and clumps are ubiquitous features of SN ejecta, likely reflecting hydrodynamical instabilities in the explosions.  Finally, the detection of H$\alpha$ and H$\beta$ in the ejecta cements the identification of SNR~0540 as a Type II SN explosion. Further analysis of the MUSE and X-shooter observations will provide new insights into the progenitor, pulsar and environment surrounding SNR~0540.

\acknowledgments

This research was supported by the Knut and Alice Wallenberg foundation. JDL acknowledges support from a UK Research and Innovation Fellowship (MR/T020784/1). We thank Francesco Taddia for help with the MUSE observing proposal. We thank Bruno Leibundgut for helpful comments on the manuscript. 

%

\vspace{5mm}
\facilities{VLT (MUSE and X-shooter)}


\software{CUPID \citep{Berry2007},
    matplotlib \citep{Hunter2007},
	Mayavi \citep{Ramachandran2011}}



\bibliography{snr0540_3d}{}

\begin{thebibliography}{}
\expandafter\ifx\csname natexlab\endcsname\relax\def\natexlab#1{#1}\fi
\providecommand{\url}[1]{\href{#1}{#1}}
\providecommand{\dodoi}[1]{doi:~\href{http://doi.org/#1}{\nolinkurl{#1}}}
\providecommand{\doeprint}[1]{\href{http://ascl.net/#1}{\nolinkurl{http://ascl.net/#1}}}
\providecommand{\doarXiv}[1]{\href{https://arxiv.org/abs/#1}{\nolinkurl{https://arxiv.org/abs/#1}}}

\bibitem[{{Abell{\'a}n} {et~al.}(2017){Abell{\'a}n}, {Indebetouw}, {Marcaide},
  {Gabler}, {Fransson}, {Spyromilio}, {Burrows}, {Chevalier}, {Cigan},
  {Gaensler}, {Gomez}, {Janka}, {Kirshner}, {Larsson}, {Lundqvist}, {Matsuura},
  {McCray}, {Ng}, {Park}, {Roche}, {Staveley-Smith}, {van Loon}, {Wheeler}, \&
  {Woosley}}]{Abellan2017}
{Abell{\'a}n}, F.~J., {Indebetouw}, R., {Marcaide}, J.~M., {et~al.} 2017,
  \apjl, 842, L24, \dodoi{10.3847/2041-8213/aa784c}

\bibitem[{{Bacon} {et~al.}(2010){Bacon}, {Accardo}, {Adjali}, {Anwand},
  {Bauer}, {Biswas}, {Blaizot}, {Boudon}, {Brau-Nogue}, {Brinchmann},
  {Caillier}, {Capoani}, {Carollo}, {Contini}, {Couderc}, {Daguis{\'e}},
  {Deiries}, {Delabre}, {Dreizler}, {Dubois}, {Dupieux}, {Dupuy}, {Emsellem},
  {Fechner}, {Fleischmann}, {Fran{\c{c}}ois}, {Gallou}, {Gharsa}, {Glindemann},
  {Gojak}, {Guiderdoni}, {Hansali}, {Hahn}, {Jarno}, {Kelz}, {Koehler},
  {Kosmalski}, {Laurent}, {Le Floch}, {Lilly}, {Lizon}, {Loupias}, {Manescau},
  {Monstein}, {Nicklas}, {Olaya}, {Pares}, {Pasquini}, {P{\'e}contal-Rousset},
  {Pell{\'o}}, {Petit}, {Popow}, {Reiss}, {Remillieux}, {Renault}, {Roth},
  {Rupprecht}, {Serre}, {Schaye}, {Soucail}, {Steinmetz}, {Streicher}, {Stuik},
  {Valentin}, {Vernet}, {Weilbacher}, {Wisotzki}, \& {Yerle}}]{Bacon2010}
{Bacon}, R., {Accardo}, M., {Adjali}, L., {et~al.} 2010, in Society of
  Photo-Optical Instrumentation Engineers (SPIE) Conference Series, Vol. 7735,
  {Ground-based and Airborne Instrumentation for Astronomy III}, ed. I.~S.
  {McLean}, S.~K. {Ramsay}, \& H.~{Takami}, 773508, \dodoi{10.1117/12.856027}

\bibitem[{{Banovetz} {et~al.}(2021){Banovetz}, {Milisavljevic}, {Sravan},
  {Fesen}, {Patnaude}, {Plucinsky}, {Blair}, {Weil}, {Morse}, {Margutti}, \&
  {Drout}}]{Banovetz2021}
{Banovetz}, J., {Milisavljevic}, D., {Sravan}, N., {et~al.} 2021, \apj, 912,
  33, \dodoi{10.3847/1538-4357/abe2a7}

\bibitem[{{Basko}(1994)}]{Basko1994}
{Basko}, M. 1994, \apj, 425, 264, \dodoi{10.1086/173983}

\bibitem[{{Berry}(2015)}]{Berry2015}
{Berry}, D.~S. 2015, Astronomy and Computing, 10, 22,
  \dodoi{10.1016/j.ascom.2014.11.004}

\bibitem[{{Berry} {et~al.}(2007){Berry}, {Reinhold}, {Jenness}, \&
  {Economou}}]{Berry2007}
{Berry}, D.~S., {Reinhold}, K., {Jenness}, T., \& {Economou}, F. 2007, in
  Astronomical Society of the Pacific Conference Series, Vol. 376, Astronomical
  Data Analysis Software and Systems XVI, ed. R.~A. {Shaw}, F.~{Hill}, \& D.~J.
  {Bell}, 425

\bibitem[{{Blondin} {et~al.}(2001){Blondin}, {Borkowski}, \&
  {Reynolds}}]{Blondin2001}
{Blondin}, J.~M., {Borkowski}, K.~J., \& {Reynolds}, S.~P. 2001, \apj, 557,
  782, \dodoi{10.1086/321674}

\bibitem[{{Brantseg} {et~al.}(2014){Brantseg}, {McEntaffer}, {Bozzetto},
  {Filipovic}, \& {Grieves}}]{Brantseg2014}
{Brantseg}, T., {McEntaffer}, R.~L., {Bozzetto}, L.~M., {Filipovic}, M., \&
  {Grieves}, N. 2014, \apj, 780, 50, \dodoi{10.1088/0004-637X/780/1/50}

\bibitem[{{Chevalier}(2006)}]{Chevalier2006}
{Chevalier}, R.~A. 2006, arXiv e-prints, astro.
\newblock \doarXiv{astro-ph/0607422}

\bibitem[{{Chugai} \& {Chevalier}(2006)}]{Chugai2006}
{Chugai}, N.~N., \& {Chevalier}, R.~A. 2006, \apj, 641, 1051,
  \dodoi{10.1086/500539}

\bibitem[{{DeLaney} {et~al.}(2010){DeLaney}, {Rudnick}, {Stage}, {Smith},
  {Isensee}, {Rho}, {Allen}, {Gomez}, {Kozasa}, {Reach}, {Davis}, \&
  {Houck}}]{DeLaney2010}
{DeLaney}, T., {Rudnick}, L., {Stage}, M.~D., {et~al.} 2010, \apj, 725, 2038,
  \dodoi{10.1088/0004-637X/725/2/2038}

\bibitem[{{Dopita} {et~al.}(1981){Dopita}, {Tuohy}, \&
  {Mathewson}}]{Dopita1981}
{Dopita}, M.~A., {Tuohy}, I.~R., \& {Mathewson}, D.~S. 1981, \apjl, 248, L105,
  \dodoi{10.1086/183635}

\bibitem[{{Eriksen} {et~al.}(2001){Eriksen}, {Morse}, {Kirshner}, \&
  {Winkler}}]{Eriksen2001}
{Eriksen}, K.~A., {Morse}, J.~A., {Kirshner}, R.~P., \& {Winkler}, P.~F. 2001,
  in American Institute of Physics Conference Series, Vol. 565, Young Supernova
  Remnants, ed. S.~S. {Holt} \& U.~{Hwang}, 193--196, \dodoi{10.1063/1.1377093}

\bibitem[{{Gabler} {et~al.}(2021){Gabler}, {Wongwathanarat}, \&
  {Janka}}]{Gabler2021}
{Gabler}, M., {Wongwathanarat}, A., \& {Janka}, H.-T. 2021, \mnras, 502, 3264,
  \dodoi{10.1093/mnras/stab116}

\bibitem[{{Ghavamian} {et~al.}(2005){Ghavamian}, {Hughes}, \&
  {Williams}}]{Ghavamian2005}
{Ghavamian}, P., {Hughes}, J.~P., \& {Williams}, T.~B. 2005, \apj, 635, 365,
  \dodoi{10.1086/497283}

\bibitem[{{Gotthelf} \& {Wang}(2000)}]{Gotthelf2000}
{Gotthelf}, E.~V., \& {Wang}, Q.~D. 2000, \apjl, 532, L117,
  \dodoi{10.1086/312568}

\bibitem[{{Hirai} {et~al.}(2018){Hirai}, {Podsiadlowski}, \&
  {Yamada}}]{Hirai2018}
{Hirai}, R., {Podsiadlowski}, P., \& {Yamada}, S. 2018, \apj, 864, 119,
  \dodoi{10.3847/1538-4357/aad6a0}

\bibitem[{{Hughes} {et~al.}(2000){Hughes}, {Rakowski}, {Burrows}, \&
  {Slane}}]{Hughes2000}
{Hughes}, J.~P., {Rakowski}, C.~E., {Burrows}, D.~N., \& {Slane}, P.~O. 2000,
  \apjl, 528, L109, \dodoi{10.1086/312438}

\bibitem[{{Hunter}(2007)}]{Hunter2007}
{Hunter}, J.~D. 2007, Computing in Science and Engineering, 9, 90,
  \dodoi{10.1109/MCSE.2007.55}

\bibitem[{{Hwang} {et~al.}(2001){Hwang}, {Petre}, {Holt}, \&
  {Szymkowiak}}]{Hwang2001}
{Hwang}, U., {Petre}, R., {Holt}, S.~S., \& {Szymkowiak}, A.~E. 2001, \apj,
  560, 742, \dodoi{10.1086/322962}

\bibitem[{{Isensee} {et~al.}(2010){Isensee}, {Rudnick}, {DeLaney}, {Smith},
  {Rho}, {Reach}, {Kozasa}, \& {Gomez}}]{Isensee2010}
{Isensee}, K., {Rudnick}, L., {DeLaney}, T., {et~al.} 2010, \apj, 725, 2059,
  \dodoi{10.1088/0004-637X/725/2/2059}

\bibitem[{{Kirshner} {et~al.}(1989){Kirshner}, {Morse}, {Winkler}, \&
  {Blair}}]{Kirshner1989}
{Kirshner}, R.~P., {Morse}, J.~A., {Winkler}, P.~F., \& {Blair}, W.~P. 1989,
  \apj, 342, 260, \dodoi{10.1086/167590}

\bibitem[{{Kozma} \& {Fransson}(1998)}]{Kozma1998}
{Kozma}, C., \& {Fransson}, C. 1998, \apj, 497, 431, \dodoi{10.1086/305452}

\bibitem[{{Krause} {et~al.}(2008){Krause}, {Birkmann}, {Usuda}, {Hattori},
  {Goto}, {Rieke}, \& {Misselt}}]{Krause2008}
{Krause}, O., {Birkmann}, S.~M., {Usuda}, T., {et~al.} 2008, Science, 320,
  1195, \dodoi{10.1126/science.1155788}

\bibitem[{{Larsson} {et~al.}(2019){Larsson}, {Spyromilio}, {Fransson},
  {Indebetouw}, {Matsuura}, {Abell{\'a}n}, {Cigan}, {Gomez}, \&
  {Leibundgut}}]{Larsson2019}
{Larsson}, J., {Spyromilio}, J., {Fransson}, C., {et~al.} 2019, \apj, 873, 15,
  \dodoi{10.3847/1538-4357/ab03d1}

\bibitem[{{Law} {et~al.}(2020){Law}, {Milisavljevic}, {Patnaude}, {Plucinsky},
  {Gladders}, {Schmidt}, {Sravan}, {Banovetz}, {Sano}, {McGraw}, {Takahashi},
  \& {Orlando}}]{Law2020}
{Law}, C.~J., {Milisavljevic}, D., {Patnaude}, D.~J., {et~al.} 2020, \apj, 894,
  73, \dodoi{10.3847/1538-4357/ab873a}

\bibitem[{{Li} {et~al.}(1993){Li}, {McCray}, \& {Sunyaev}}]{Li1993}
{Li}, H., {McCray}, R., \& {Sunyaev}, R.~A. 1993, \apj, 419, 824,
  \dodoi{10.1086/173534}

\bibitem[{{Liu} {et~al.}(2015){Liu}, {Tauris}, {R{\"o}pke}, {Moriya},
  {Kruckow}, {Stancliffe}, \& {Izzard}}]{Liu2015}
{Liu}, Z.-W., {Tauris}, T.~M., {R{\"o}pke}, F.~K., {et~al.} 2015, \aap, 584,
  A11, \dodoi{10.1051/0004-6361/201526757}

\bibitem[{{Lundqvist} {et~al.}(2011){Lundqvist}, {Lundqvist}, {Bj{\"o}rnsson},
  {Olofsson}, {Pires}, {Shibanov}, \& {Zyuzin}}]{Lundqvist2011}
{Lundqvist}, N., {Lundqvist}, P., {Bj{\"o}rnsson}, C.~I., {et~al.} 2011,
  \mnras, 413, 611, \dodoi{10.1111/j.1365-2966.2010.18159.x}

\bibitem[{{Lundqvist} {et~al.}(2021){Lundqvist}, {Lundqvist}, \&
  {Shibanov}}]{Lundqvist2021}
{Lundqvist}, P., {Lundqvist}, N., \& {Shibanov}, Y.~A. 2021, arXiv e-prints,
  arXiv:2109.03287.
\newblock \doarXiv{2109.03287}

\bibitem[{{Lundqvist} {et~al.}(2020){Lundqvist}, {Lundqvist}, {Vlahakis},
  {Bj{\"o}rnsson}, {Dickel}, {Matsuura}, {Shibanov}, {Zyuzin}, \&
  {Olofsson}}]{Lundqvist2020}
{Lundqvist}, P., {Lundqvist}, N., {Vlahakis}, C., {et~al.} 2020, \mnras, 496,
  1834, \dodoi{10.1093/mnras/staa1675}

\bibitem[{{Manchester} {et~al.}(1993){Manchester}, {Staveley-Smith}, \&
  {Kesteven}}]{Manchester1993}
{Manchester}, R.~N., {Staveley-Smith}, L., \& {Kesteven}, M.~J. 1993, \apj,
  411, 756, \dodoi{10.1086/172877}

\bibitem[{{Martin} {et~al.}(2021){Martin}, {Milisavljevic}, \&
  {Drissen}}]{Martin2021}
{Martin}, T., {Milisavljevic}, D., \& {Drissen}, L. 2021, \mnras, 502, 1864,
  \dodoi{10.1093/mnras/staa4046}

\bibitem[{{Mignani} {et~al.}(2010){Mignani}, {Sartori}, {de Luca}, {Rudak},
  {S{\l}owikowska}, {Kanbach}, \& {Caraveo}}]{Mignani2010}
{Mignani}, R.~P., {Sartori}, A., {de Luca}, A., {et~al.} 2010, \aap, 515, A110,
  \dodoi{10.1051/0004-6361/200913870}

\bibitem[{{Milisavljevic} \& {Fesen}(2013)}]{Milisavljevic2013}
{Milisavljevic}, D., \& {Fesen}, R.~A. 2013, \apj, 772, 134,
  \dodoi{10.1088/0004-637X/772/2/134}

\bibitem[{{Milisavljevic} \& {Fesen}(2015)}]{Milisavljevic2015}
---. 2015, Science, 347, 526, \dodoi{10.1126/science.1261949}

\bibitem[{{Milisavljevic} \& {Fesen}(2017)}]{Milisavljevic2017}
---. 2017, {The Supernova - Supernova Remnant Connection}, ed. A.~W. {Alsabti}
  \& P.~{Murdin}, 2211, \dodoi{10.1007/978-3-319-21846-5\_97}

\bibitem[{{Milisavljevic} {et~al.}(2015){Milisavljevic}, {Margutti}, {Kamble},
  {Patnaude}, {Raymond}, {Eldridge}, {Fong}, {Bietenholz}, {Challis},
  {Chornock}, {Drout}, {Fransson}, {Fesen}, {Grindlay}, {Kirshner}, {Lunnan},
  {Mackey}, {Miller}, {Parrent}, {Sanders}, {Soderberg}, \&
  {Zauderer}}]{Milisavljevic2015_14C}
{Milisavljevic}, D., {Margutti}, R., {Kamble}, A., {et~al.} 2015, \apj, 815,
  120, \dodoi{10.1088/0004-637X/815/2/120}

\bibitem[{{Morse} {et~al.}(2006){Morse}, {Smith}, {Blair}, {Kirshner},
  {Winkler}, \& {Hughes}}]{Morse2006}
{Morse}, J.~A., {Smith}, N., {Blair}, W.~P., {et~al.} 2006, \apj, 644, 188,
  \dodoi{10.1086/503313}

\bibitem[{{Orlando} {et~al.}(2021){Orlando}, {Wongwathanarat}, {Janka},
  {Miceli}, {Ono}, {Nagataki}, {Bocchino}, \& {Peres}}]{Orlando2021}
{Orlando}, S., {Wongwathanarat}, A., {Janka}, H.~T., {et~al.} 2021, \aap, 645,
  A66, \dodoi{10.1051/0004-6361/202039335}

\bibitem[{{Osterbrock}(1989)}]{Osterbrock1989}
{Osterbrock}, D.~E. 1989, {Astrophysics of gaseous nebulae and active galactic
  nuclei}

\bibitem[{{Pietrzy{\'n}ski} {et~al.}(2019){Pietrzy{\'n}ski}, {Graczyk},
  {Gallenne}, {Gieren}, {Thompson}, {Pilecki}, {Karczmarek}, {G{\'o}rski},
  {Suchomska}, {Taormina}, {Zgirski}, {Wielg{\'o}rski}, {Ko{\l}aczkowski},
  {Konorski}, {Villanova}, {Nardetto}, {Kervella}, {Bresolin}, {Kudritzki},
  {Storm}, {Smolec}, \& {Narloch}}]{Pietrzynski2019}
{Pietrzy{\'n}ski}, G., {Graczyk}, D., {Gallenne}, A., {et~al.} 2019, \nat, 567,
  200, \dodoi{10.1038/s41586-019-0999-4}

\bibitem[{{Ramachandran} \& {Varoquaux}(2011)}]{Ramachandran2011}
{Ramachandran}, P., \& {Varoquaux}, G. 2011, Computing in Science and
  Engineering, 13, 40, \dodoi{10.1109/MCSE.2011.35}

\bibitem[{{Rest} {et~al.}(2011){Rest}, {Foley}, {Sinnott}, {Welch}, {Badenes},
  {Filippenko}, {Bergmann}, {Bhatti}, {Blondin}, {Challis}, {Damke}, {Finley},
  {Huber}, {Kasen}, {Kirshner}, {Matheson}, {Mazzali}, {Minniti}, {Nakajima},
  {Narayan}, {Olsen}, {Sauer}, {Smith}, \& {Suntzeff}}]{Rest2011}
{Rest}, A., {Foley}, R.~J., {Sinnott}, B., {et~al.} 2011, \apj, 732, 3,
  \dodoi{10.1088/0004-637X/732/1/3}

\bibitem[{{Rimoldi} {et~al.}(2016){Rimoldi}, {Portegies Zwart}, \&
  {Rossi}}]{Rimoldi2016}
{Rimoldi}, A., {Portegies Zwart}, S., \& {Rossi}, E.~M. 2016, Computational
  Astrophysics and Cosmology, 3, 2, \dodoi{10.1186/s40668-016-0015-4}

\bibitem[{{Sandin} {et~al.}(2013){Sandin}, {Lundqvist}, {Lundqvist},
  {Bj{\"o}rnsson}, {Olofsson}, \& {Shibanov}}]{Sandin2013}
{Sandin}, C., {Lundqvist}, P., {Lundqvist}, N., {et~al.} 2013, \mnras, 432,
  2854, \dodoi{10.1093/mnras/stt641}

\bibitem[{{Sandoval} {et~al.}(2021){Sandoval}, {Hix}, {Messer}, {Lentz}, \&
  {Harris}}]{Sandoval2021}
{Sandoval}, M.~A., {Hix}, W.~R., {Messer}, O.~E.~B., {Lentz}, E.~J., \&
  {Harris}, J.~A. 2021, arXiv e-prints, arXiv:2106.01389.
\newblock \doarXiv{2106.01389}

\bibitem[{{Seitenzahl} {et~al.}(2018){Seitenzahl}, {Vogt}, {Terry},
  {Ghavamian}, {Dopita}, {Ruiter}, \& {Sukhbold}}]{Seitenzahl2018}
{Seitenzahl}, I.~R., {Vogt}, F. P.~A., {Terry}, J.~P., {et~al.} 2018, \apjl,
  853, L32, \dodoi{10.3847/2041-8213/aaa958}

\bibitem[{{Serafimovich} {et~al.}(2005){Serafimovich}, {Lundqvist}, {Shibanov},
  \& {Sollerman}}]{Serafimovich2005}
{Serafimovich}, N.~I., {Lundqvist}, P., {Shibanov}, Y.~A., \& {Sollerman}, J.
  2005, Advances in Space Research, 35, 1106, \dodoi{10.1016/j.asr.2005.01.071}

\bibitem[{{Seward} {et~al.}(1984){Seward}, {Harnden}, \&
  {Helfand}}]{Seward1984}
{Seward}, F.~D., {Harnden}, F.~R., J., \& {Helfand}, D.~J. 1984, \apjl, 287,
  L19, \dodoi{10.1086/184388}

\bibitem[{{Smith}(2008)}]{Smith2008}
{Smith}, N. 2008, \nat, 455, 201, \dodoi{10.1038/nature07269}

\bibitem[{{Smith} {et~al.}(2001){Smith}, {Humphreys}, {Davidson}, {Gehrz},
  {Schuster}, \& {Krautter}}]{Smith2001}
{Smith}, N., {Humphreys}, R.~M., {Davidson}, K., {et~al.} 2001, \aj, 121, 1111,
  \dodoi{10.1086/318748}

\bibitem[{{Smith} {et~al.}(2011){Smith}, {Li}, {Silverman}, {Ganeshalingam}, \&
  {Filippenko}}]{Smith2011}
{Smith}, N., {Li}, W., {Silverman}, J.~M., {Ganeshalingam}, M., \&
  {Filippenko}, A.~V. 2011, \mnras, 415, 773,
  \dodoi{10.1111/j.1365-2966.2011.18763.x}

\bibitem[{{Sollerman} {et~al.}(2020){Sollerman}, {Fransson}, {Barbarino},
  {Fremling}, {Horesh}, {Kool}, {Schulze}, {Sfaradi}, {Yang}, {Bellm},
  {Burruss}, {Cunningham}, {De}, {Drake}, {Golkhou}, {Green}, {Kasliwal},
  {Kulkarni}, {Kupfer}, {Laher}, {Masci}, {Rodriguez}, {Rusholme}, {Williams},
  {Yan}, \& {Zolkower}}]{Sollerman2020}
{Sollerman}, J., {Fransson}, C., {Barbarino}, C., {et~al.} 2020, \aap, 643,
  A79, \dodoi{10.1051/0004-6361/202038960}

\bibitem[{{Stockinger} {et~al.}(2020){Stockinger}, {Janka}, {Kresse}, {Melson},
  {Ertl}, {Gabler}, {Gessner}, {Wongwathanarat}, {Tolstov}, {Leung}, {Nomoto},
  \& {Heger}}]{Stockinger2020}
{Stockinger}, G., {Janka}, H.~T., {Kresse}, D., {et~al.} 2020, \mnras, 496,
  2039, \dodoi{10.1093/mnras/staa1691}

\bibitem[{{Utrobin} {et~al.}(2019){Utrobin}, {Wongwathanarat}, {Janka},
  {M{\"u}ller}, {Ertl}, \& {Woosley}}]{Utrobin2019}
{Utrobin}, V.~P., {Wongwathanarat}, A., {Janka}, H.~T., {et~al.} 2019, \aap,
  624, A116, \dodoi{10.1051/0004-6361/201834976}

\bibitem[{{Williams} {et~al.}(2008){Williams}, {Borkowski}, {Reynolds},
  {Raymond}, {Long}, {Morse}, {Blair}, {Ghavamian}, {Sankrit}, {Hendrick},
  {Smith}, {Points}, \& {Winkler}}]{Williams2008}
{Williams}, B.~J., {Borkowski}, K.~J., {Reynolds}, S.~P., {et~al.} 2008, \apj,
  687, 1054, \dodoi{10.1086/592139}

\bibitem[{{Wongwathanarat} {et~al.}(2015){Wongwathanarat}, {M{\"u}ller}, \&
  {Janka}}]{Wongwathanarat2015}
{Wongwathanarat}, A., {M{\"u}ller}, E., \& {Janka}, H.~T. 2015, \aap, 577, A48,
  \dodoi{10.1051/0004-6361/201425025}

\end{thebibliography}
\bibliographystyle{aasjournal}

\end{document}